\documentclass[a4paper,fleqn,usenatbib]{mnras}

\usepackage[T1]{fontenc}
\usepackage{ae,aecompl}

\usepackage{graphicx}	% Including figure files
\usepackage{amsmath}	% Advanced maths commands
\usepackage{amssymb}	% Extra maths symbols
\usepackage{xspace}

\usepackage{natbib,bm}
\usepackage[usenames]{color}
\usepackage{textcomp}

\newcommand{\bd}{\begin{displaymath}}
\newcommand{\ed}{\end{displaymath}}
\newcommand{\be}{\begin{equation}}
\newcommand{\ee}{\end{equation}}
\newcommand{\beaa}{\begin{eqnarray*}}
\newcommand{\eeaa}{\end{eqnarray*}}
\newcommand{\bea}{\begin{eqnarray}}
\newcommand{\eea}{\end{eqnarray}}

\def\hequad{HE\,0435$-$1223}

\def\wfilens{WFI2033$-$4723}
\def\blens{B1608$+$656}
\def\rxjlens{RXJ1131$-$1231}

\def\PaperI{H0LiCOW~Paper~I\xspace}
\def\PaperII{H0LiCOW~Paper~II\xspace}
\def\PaperIII{H0LiCOW~Paper~III\xspace}
\def\PaperIV{H0LiCOW~Paper~IV\xspace}

\def\Ok{\Omega_{\rm k}}
\def\Ode{\Omega_{\rm DE}}
\def\Om{\Omega_{\rm m}}
\def\OL{\Omega_{\Lambda}}
\def\Neff{N_{\rm eff}}

\def\zd{z_{\rm d}}
\def\zs{z_{\rm s}}

\def\hst{\textit{HST}}

\def\GLEE{{\sc Glee}\xspace}

% ...
\usepackage{soul} % This is for \st{}
\definecolor{darkgreen}{rgb}{0.0, 0.5, 0.0}

\newcommand{\specialcell}[2][c]{%
  \begin{tabular}[#1]{@{}c@{}}#2\end{tabular}}

\def\Ode{\Omega_{\rm de}}

\def\hc{$H_0$\xspace}
\def\nnu{${\rm N_{eff}}$\xspace}
\def\mnu{${\rm \Sigma m_{\nu}}$\xspace}

\def\LCDM{$\Lambda {\rm CDM}$\xspace}
\def\oLCDM{${\rm o}\Lambda {\rm CDM}$\xspace}
\def\wCDM{${\rm {\it w}CDM}$\xspace}
\def\UHO{${\rm U}H_0$\xspace}
\def\ULCDM{${\rm U \Lambda CDM}$\xspace}
\def\UoLCDM{${\rm Uo}\Lambda {\rm CDM}$\xspace}
\def\UwCDM{${\rm U{\it w}CDM}$\xspace}

\def\owCDM{${\rm o{\it w}CDM}$\xspace}
\def\mnuLCDM{${\rm m_{\nu}}\Lambda {\rm CDM}$\xspace}
\def\nnuLCDM{${\rm N_{eff}}\Lambda {\rm CDM}$\xspace}
\def\nnumnuLCDM{${\rm N_{eff}m_{\nu}}\Lambda {\rm CDM}$\xspace}

\title[HE0435 time delays and \hc to 3.8\%]{H0LiCOW V. New COSMOGRAIL
time delays of \hequad: \hc
to 3.8\% precision from strong lensing in a flat \LCDM model}

\author[V.~Bonvin et al.]{
V.~Bonvin,$^{1}$\thanks{E-mail: vivien.bonvin@epfl.ch}
F.~Courbin,$^{1}$
S.~H.~Suyu,$^{2, 3, 4}$
P.~J.~Marshall,$^{5}$
C.~E.~Rusu,$^{6}$
D.~Sluse,$^{7}$\newauthor
\ M.~Tewes,$^{8}$
K.~C.~Wong,$^{9, 4}$
T.~Collett,$^{10}$
C.~D.~Fassnacht,$^{7}$
T.~Treu,$^{11}$\newauthor
\ M.~W.~Auger,$^{12}$
S.~Hilbert,$^{13, 14}$
L.~V.~E.~Koopmans,$^{15}$
G.~Meylan,$^{1}$\newauthor
\ N.~Rumbaugh,$^{11}$
A.~Sonnenfeld,$^{16,11,17}$
and C.~Spiniello$^{2}$\\
$^{1}$Institute of Physics, Laboratory of Astrophysics, Ecole 
Polytechnique F\'ed\'erale de Lausanne (EPFL),\\ \ Observatoire 
de Sauverny, 1290 Versoix, Switzerland.\\
$^{2}$Max Planck Institute for Astrophysics, Karl-Schwarzschild-Strasse
1, D-85740 Garching, Germany\\
$^{3}$Physik-Department, Technische Universit\"at M\"unchen, 
James-Franck-Stra\ss{}e~1, 85748 Garching, Germany\\
$^{4}$Institute of Astronomy and Astrophysics, Academia Sinica,
P.O.~Box
23-141, Taipei 10617, Taiwan\\
$^{5}$Kavli Institute for Particle Astrophysics and Cosmology, Stanford
University, 452 Lomita Mall, Stanford, CA 94035, USA\\
$^{6}$ Department of Physics, University of California, Davis, CA 95616,
USA\\
$^{7}$STAR Institute, Quartier Agora - All\'ee du six Ao\^ut, 19c B-4000
Li\`ege, Belgium\\
$^{8}$Argelander-Institut f\"ur Astronomie, Auf dem H\"ugel 71, D-53121
Bonn, Germany\\
$^{9}$National Astronomical Observatory of Japan, 2-21-1 Osawa, Mitaka,
Tokyo 181-8588, Japan\\
$^{10}$Institute of Cosmology and Gravitation, University of Portsmouth,
Burnaby Rd, Portsmouth PO1 3FX, UK\\
$^{11}$Department of Physics and Astronomy, University of California,
Los Angeles, CA 90095, USA \\
$^{12}$Institute of Astronomy, University of Cambridge, Madingley Road,
Cambridge CB3 0HA, UK \\
$^{13}$ Exzellenzcluster Universe, Boltzmannstr. 2, 85748 Garching,
Germany\\
$^{14}$ Ludwig-Maximilians-Universit{\"a}t,
Universit{\"a}ts-Sternwarte, Scheinerstr. 1, 81679 M{\"u}nchen,
Germany\\
$^{15}$Kapteyn Astronomical Institute, University of Groningen,
P.O. Box 800, 9700-AV Groningen, The Netherlands \\
$^{16}$Kavli IPMU (WPI), UTIAS, The University of Tokyo, Kashiwa, Chiba
277-8583, Japan\\
$^{17}$Physics Department, University of California, Santa Barbara, CA,
93106, USA
}

\pubyear{2017}

\begin{document}
\label{firstpage}
\pagerange{\pageref{firstpage}--\pageref{lastpage}}
\maketitle

%-------------------------------------------------------------------------------

\begin{abstract}

We present a new measurement of the Hubble Constant \hc\ and other
cosmological parameters based on the joint analysis of three
multiply-imaged quasar systems with measured gravitational time
delays. First, we measure the time delay of \hequad\
from 13-year light curves obtained as part of the COSMOGRAIL project.
Companion papers detail the
modeling of the main deflectors and line of sight effects, and how
these data are combined to determine the time-delay distance of
\hequad. Crucially, the measurements are carried out blindly
with respect to cosmological parameters in order to avoid confirmation
bias. We then combine the time-delay distance of
\hequad\ with previous measurements from systems \blens\ and \rxjlens\
to create a Time Delay Strong Lensing probe (TDSL). In flat \LCDM\ with
free matter and energy density, we find
\hc$=71.9^{+2.4}_{-3.0}\ {\rm km\, s^{-1}\, Mpc^{-1}}$ and
$\Omega_{\Lambda}=0.62^{+0.24}_{-0.35}$. This measurement is
completely independent of, and in agreement with, the local
distance ladder measurements of \hc. We
explore more general cosmological models combining TDSL with other
probes, illustrating its power to break degeneracies inherent to other
methods. The joint constraints from TDSL and Planck are
\hc = $69.2_{-2.2}^{+1.4}\ {\rm km\, s^{-1}\, Mpc^{-1}}$, 
$\Omega_{\Lambda}=0.70_{-0.01}^{+0.01}$ and
$\Ok=0.003_{-0.006}^{+0.004}$ in open \LCDM and
\hc$=79.0_{-4.2}^{+4.4}\ {\rm km\, s^{-1}\, Mpc^{-1}}$, 
$\Ode=0.77_{-0.03}^{+0.02}$ and
$w=-1.38_{-0.16}^{+0.14}$ in flat \wCDM. In combination with
Planck and
Baryon Acoustic Oscillation data, when relaxing the constraints on the
numbers of relativistic species we find
\nnu= $3.34_{-0.21}^{+0.21}$ in \nnuLCDM and when relaxing the total 
mass of neutrinos we find \mnu\ $\leq\, 0.182$~eV in \mnuLCDM.
Finally,
in an open \wCDM in combination with Planck and CMB lensing we find
\hc$=77.9_{-4.2}^{+5.0}\ {\rm km\, s^{-1}\, Mpc^{-1}}$, 
$\Ode=0.77_{-0.03}^{+0.03}$,
$\Ok=-0.003_{-0.004}^{+0.004}$ and
$w=-1.37_{-0.23}^{+0.18}$.
\end{abstract}


%=========================================================
%			INTRODUCTION
%=========================================================

\section{Introduction}
\label{sec:intro}

In the past decade, the Standard Cosmological Model, \LCDM, which
assumes the existence of either a cosmological constant or a form of
dark energy with equation of state $w = -1$, and large scale structure
predominantly composed of Cold Dark Matter, has been firmly established
given observations to date \citep[e.g.][]{Hinshaw2013, Planck2015}. From
a minimal set of 6 parameters describing $\Lambda$CDM, one can in
principle infer the value of other parameters such as the current
expansion rate of the Universe, $H_0$. However, such an inference
involves strong assumptions about the cosmological model, such as the
absence of curvature or a constant equation of state for the dark
energy. Conversely, we can relax these assumptions and explore models
beyond flat-\LCDM using a wider set of cosmological probes. In this
case, the analysis benefits greatly from independent measurements of
$H_0$ from observations of distance probes such as the distance ladder 
or
water masers \citep[see 
e.g.][for a review]{Treu2010, Weinberg2013, 
Treu2016}. As \citet{Weinberg2013} point
out, the Figure of Merit of any stage III or stage IV cosmological
survey improves by 40\% if an independent measurement of $H_0$ is
available to a precision of 1\%.

The ``time-delay distances'' in gravitationally lensed quasar systems
offer an opportunity to measure $H_0$ independently of any other
cosmological probe. First suggested by \citet{Refsdal1964}, this
approach involves measuring the time delays between multiple images of a
distant source produced by a foreground lensing object. The time delays
depend on the matter distribution in the lens (galaxy), on the overall
matter distribution along the line-of-sight and on the cosmological
parameters. The time delays are related to the so-called time-delay
distance $D_{\Delta t}$ to the lens and the source, which is primarily
sensitive to \hc and has a weak dependence on the matter density $\Om$, 
the dark energy density $\Ode$, the dark energy equation
of state, $w$, and on the curvature parameter $\Omega_k$
\citep[e.g.][]{Linder2011, Suyu2010b}.

The first critical step for the method to work is the measurement of the
time delays from a photometric monitoring campaign to measure the shift
in time between the light curves of the lensed images of quasars. Such
monitoring campaigns must be long enough, and have good enough temporal
sampling, to catch all possible (and usually small) photometric
variations in the light curves. This is the goal of the COSMOGRAIL
collaboration: the COSmological MOnitoring of GRAvItational Lenses,
which has been monitoring about 20 lensed quasars with 1m-class and 2m-class
telescopes since 2004 \citep[e.g.][]{Courbin2005, Eigenbrod2006a, 
Bonvin2016}.
The target precision for the time delay measurements is a few percent
or better, because the error on the time delays propagates linearly to
first order on the
cosmological distance measurement. Examples of COSMOGRAIL results 
include \citet{Courbin2011},
\citet{Tewes2013b}, \citet{Rathnakumar2013}, and \citet{Eulaers2013}.

The second critical step is the modeling of the lens galaxy. Indeed,
time-delay measurements alone can constrain only a combination of the 
time-delay
distance and the surface density of the lens around the quasar images
\citep{Kochanek2002}. Additional constraints on the density profile of
the lens are therefore required in order to convert observed time delays into
inferences of the time-delay distance. These constraints can be derived 
from velocity
dispersion measurements, and the radial magnification of the extended, lensed arc
image of the quasar host galaxy
\citep[e.g.][]{Suyu2010b, Suyu2014}. Ideal targets for this purpose are 
lensed
quasars with a prominent host, which offer strong constraints on the
density profile slope of the foreground lens.

In modeling the lens mass distribution, special care has to be paid to the mass-sheet
degeneracy (MSD), and, more generally, the source-position transformation
(SPT) \citep[e.g.][]{Falco1985, Wucknitz2002, Schneider2013,
Schneider2014, Xu2016,
Unruh2016}.
These can be seen as degeneracies in the choice of the
gravitational lensing potential that leave all the lensing observables
invariant except for the modeled time delay, $\Delta t$.
In other words, a wrong model of the main lens mass
distribution can perfectly
fit the observed morphology the lensing system, and yet result in an
inaccurate
inference of the time-delay distance. Priors and spectroscopic 
constraints
on the dynamics of the main lens therefore play a 
critical role in avoiding systematic biases.
In addition,
perturbations to the lens potential by the distribution of mass along
the line-of-sight also create degeneracies in the lens modeling. The latter can be
mitigated with a measurement of the mass
distribution along the line-of-sight, for example by using
spectroscopic redshift measurements of the galaxies in the lens
environment \citep[e.g.][]{Fassnacht2006, Wong2011}, comparisons between
galaxy number counts in the real data and in simulations
\citep{Hilbert2007, Hilbert2009, Fassnacht2011, Collett2013,
Greene2013, Suyu2013, McCully2016} or
using weak-lensing measurements (Tihhonova et al., in prep.)

The H0LiCOW collaboration (\hc Lenses in COSMOGRAIL's Wellspring)
capitalizes on the efforts of COSMOGRAIL to measure accurate time delays,
and on high quality auxiliary data from {\it Hubble Space Telescope} 
({\it HST}) and 10-m class ground-based
telescopes, to constrain cosmology. The H0LiCOW sample consists of five well-selected
targets, each with exquisite time-delay measurements. \blens,
monitored in
radio band with the VLA \citep{Fassnacht2002}, and \rxjlens, monitored
by
COSMOGRAIL in the visible \citep{Tewes2013b}, have already shown promising
results, with relative precisions on the time-delay distance of 5\%
and 6.6\% respectively
\citep[][]{Suyu2010b, Suyu2014}.

\begin{table*}
 \caption{Optical monitoring campaigns of \hequad. The sampling is the
 mean number of days between the observations, not considering the
 seasonal gaps.}
  \centering
 \begin{tabular}{l c c c r r r c r}
  \hline
  Telescope & Camera  & FoV & Pixel & Period of observation & $\#$obs & 
Exp.time & median FWHM & Sampling\\
  \hline
  Euler & C2 & 11'$\times$11' & 0.344'' & Jan 2004 - Mar 2010 & 301 & 
5$\times$360s &1.37'' & 6 days \\
  Euler & ECAM & 14.2'$\times$14.2' & 0.215'' & Sep 2010 - Mar 2016 & 
301 & 5$\times$360s &1.39'' & 4 days \\
  Mercator & MEROPE & 6.5'$\times$6.5' & 0.190'' & Sep 2004 - Dec 2008 
& 
104 & 5$\times$360s &1.59'' & 11 days \\
  Maidanak & SITE & 8.9'$\times$3.5' & 0.266'' & Oct 2004 - Jul 2006 & 
26 & 10$\times$180s &1.31'' & 16 days \\
  Maidanak & SI & 18.1'$\times$18.1' & 0.266'' & Aug 2006 - Jan 2007 & 
8 
& 6$\times$300s &1.31'' & 16 days \\
  SMARTS & ANDICAM & 10'$\times$10' & 0.300'' & Aug 2003 - Apr 2005 & 
136 & 3$\times$300s &$\leq$1.80'' & 4 days \\
  \hline
  {\bf TOTAL} & - & - & - & Aug 2003 - Mar 2016 & 876 & 394.5h & - & 
3.6 
days \\
  \hline
 \end{tabular}
 \label{tab:monitoring}
\end{table*}

\begin{figure*}
  \centering
    \includegraphics[width=0.99\textwidth]{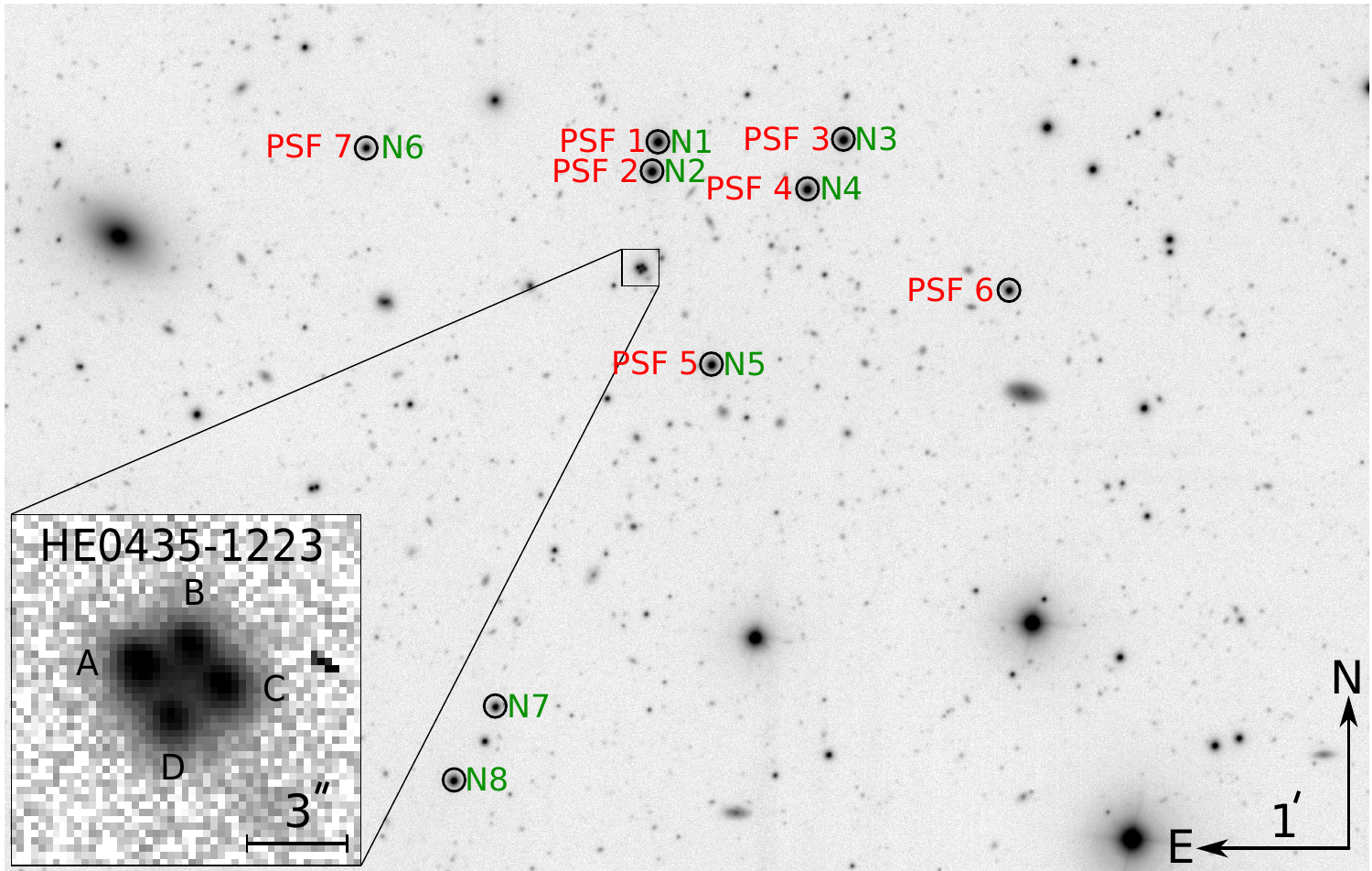}
    \caption{Part of the field of view of EulerCAM installed on the
    Swiss 1.2m telescope around the quasar \hequad. This image is a
    combination of 100 exposures of 360s each, for a total exposure time
    of 10 hours. The stars used to build a PSF model for each
    EulerCAM exposure are circled and labeled PSF1 to PSF7 in red, and
    the stars used for the photometric calibrations are circled and 
labeled
    N1 to N8 in green. The insert in the bottom left shows the single,
    360s exposure of the lens, for reference. Note that
    photometric and spectroscopic redshifts are available for many
    galaxies in the field of view (see \PaperII and
    \PaperIII for details). }
    \label{fig:he0435_deepfield}
\end{figure*}

This paper is part of the H0LiCOW series, focusing on the quadruple
lensed
quasar \hequad\
($\alpha$(2000):~04h~38m~14.9s; $\delta$(2000):~-12$^\circ$17\arcmin14\farcs4)
\citep{Wisotzki2000, Wisotzki2002}
discovered during the Hamburg/ESO Survey (HES) for bright quasars in the
Southern Hemisphere. The source redshift has been measured by
\citet{Sluse2012} as $\zs = 1.693$, and the redshift of the lens has been
measured by \citet{Morgan2005} and \citet{Eigenbrod2006b} as $\zd =
0.4546 \pm 0.0002$. The lens lies in a group of galaxies of at least 12
members. A first measurement of the time delay for \hequad\ was presented
in \citet{Courbin2011}. In this work, we present a
significant improvement of the time delay measurement, with twice as
long light curves as in \citet{Courbin2011}. The other H0LiCOW papers
include an
overview of the project (Suyu et al., submitted; hereafter \PaperI),
a spectroscopic survey of the field of \hequad\ and a
characterization of the groups along the line-of-sight (Sluse et al.,
submitted; hereafter \PaperII), a photometric survey of the
field of \hequad\ with an estimate of the effect of the external
line-of-sight structure (Rusu et al., submitted; hereafter \PaperIII),
and a detailed modeling of the lens and the inference of the
time-delay distance along with cosmological results for \hequad\ (Wong
et al., submitted; hereafter \PaperIV). In the present paper we
combine the results for \hequad\ with those from the other two
lensed quasars already published, and with other cosmological
datasets \citep{Bennett2013, Hinshaw2013, Planck2015}.

This paper is organized as follows. Section~\ref{sec:data} presents the
COSMOGRAIL optical monitoring data and its reduction process. Section~\ref{sec:timedelay}
presents the time-delay measurements and related uncertainties.
Section~\ref{sec:modeling} summarizes the main steps of the 
field-of-view analysis detailed in \PaperII and \PaperIII and the lens
modeling detailed in \PaperIV that lead to the time-delay
distance determination. Section~\ref{sec:cosmography} combines the
time-delay distance of \hequad\ and other lenses, and with additional 
cosmological
datasets, in order to make the best possible inferences
of cosmological parameters. Finally,
Section~\ref{sec:conclusion} presents our conclusions and future
prospects in the light of these results.

%=========================================================
%			DATA ACQUISITION
%=========================================================

\section{Photometric monitoring data}
\label{sec:data}

\hequad\ has been monitored since 2003 as part of the COSMOGRAIL
program and in collaboration with the \citet{Kochanek2006} team. The
data acquired from autumn 2003 to spring 2010 were
presented in \citet{Courbin2011}. Here, we double the monitoring period, adding
observations taken between autumn 2010 and spring 2016. Our monitoring
sites include two Northern telescopes: the 1.2m Belgian Mercator
telescope located at the Roque de Los Muchachos Observatory, La Palma,
Canary Islands (Spain) and the 1.5m telescope located at the Maidanak
Observatory (Uzbekistan). The average observing cadence was 11 and 16
days respectively at these sites. These telescopes ceased taking data for
COSMOGRAIL in December 2008. In the Southern hemisphere, the Swiss 1.2m
Euler telescope located at the ESO La Silla observatory (Chile) has monitored
\hequad\ since 2004. Two cameras were used: the C2 and the EulerCAM
instruments, with an average cadence of 6 days and 4 days respectively.
We also make use of the data obtained at the 1.3m SMARTS ANDICAM camera
at Cerro Tololo Inter-American Observatory. Note that we do not
re-analyse the SMARTS data, but use directly the published photometric
measurements \citep{Kochanek2006}. Table~\ref{tab:monitoring} gives a
detailed summary of the observations.

\subsection{Data reduction}

The full data set consists of two distinct blocks that do not overlap in
time and that we treat independently. The first block includes the
Mercator, Maidanak and Euler-C2 data, to which we add the published
SMARTS photometry. The detailed processing  and the relative photometric
calibration of these curves is presented in $\mathrm{Section~\,2.2}$ of
\citet{Courbin2011}. The second block consists of the 301 new data points obtained with
EulerCAM that we reduce with the pipeline described in
$\mathrm{Section~\,3}$ of \citet{Tewes2013b}, whose main steps can be
summarized as follows:

\begin{enumerate}

\item Each image is corrected for bias and readout effects. We then
apply a flat-field correction using a high signal-to-noise master
sky-flat which we correct for a pattern generated by the shutter opening
and closing times. A spatially variable sky background frame is then
constructed using the SExtractor software \citep{Bertin1996} and we
subtract it from the data frame. All the frames are aligned and analyzed
to carry out the photometric measurements.
Figure~\ref{fig:he0435_deepfield} presents a stack of the 100 EulerCAM
images with a seeing smaller than 1.14 arcsec.

\item The photometric measurements of the four blended images of
\hequad\ are obtained using deconvolution photometry using the MCS
deconvolution algorithm \citep[][]{Magain1998, Cantale2016a}. To do
this, the Point Spread Function (PSF) is measured, for each exposure
individually, using the seven stars labeled PSF1 to PSF7 on
Figure~\ref{fig:he0435_deepfield}. A simultaneous deconvolution of all
the frames is then carried out, leading to a model composed of a deep
image representing extended sources, and a catalog of point sources with
improved resolution and sampling. During the deconvolution process the
data are decomposed into a sum of analytical point sources (the quasar
images) and of a numerical pixel channel containing the image of the
lensing galaxy and of any potential extended object.

\item We compute a multiplicative median normalization coefficient for
each exposure, using several deconvolved reference stars. If possible at
all, we use stars whose color are similar to that of the quasar. In the
case of \hequad, we use 8 reference stars, labeled N1 to N8 in
Figure~\ref{fig:he0435_deepfield}. We then apply the normalization
coefficient to the deconvolved images of the point sources. Their
intensities are returned for every frame, hence leading to the light
curves.

\end{enumerate}

The upper panel of Figure~\ref{fig:he0435_lightcurve} presents the
13-year-long COSMOGRAIL light curves of \hequad, including the data from
\citet{Courbin2011} and our new data. The similarity between the 4 light curves is
immediately noticeable. However, it can also be noted that they would
not superpose perfectly when shifted in time and magnitude, due to
``extrinsic variability'' which is interpreted as being caused by
microlensing by stars in the lensing galaxy \citep[see
e.g.][]{Blackburne2014, Braibant2014}. These extrinsic contributions 
are clearly seen
here on timescales from a few weeks to several years, in the form of an
evolution of the magnitude-separation between the light curves. They
must be handled properly in order to measure time delays with high
accuracy.

\subsection{On the importance of long light curves}

Given the limited photometric precision of the COSMOGRAIL images, long-term
monitoring is crucial to the time delay measurement, for two main
reasons. First, one needs to catch enough intrinsic photometric
variations in the quasar light curves in order to identify common
structures. In the present case, these can be found
on average 2-3 times per observing season, with some seasons displaying
more prominent structures than others. Inflexion points in the light 
curves are
most precious to constrain the time delays. For example, dips and peaks
with an amplitude of nearly half a magnitude can be observed in several
seasons: 2004-2005, 2012-2013 and 2015-2016. Second, the extrinsic
variability related to microlensing must be taken into account (e.g. by
modeling and removing it) to avoid time-delay measurement biases. Any
simple and well-constrainable model is likely not sufficient to capture
all aspects of this extrinsic variability, and might result in residual
biases. The availability of decade long light curves allows us to check
for potentially significant biases by analysing subsets of the full
data, and certainly to reduce residual ones.

%=========================================================
%			TIME DELAY MEASUREMENT
%=========================================================

\section{Time-delay measurement}
\label{sec:timedelay}

With the light curves in hand, the time delays can be measured using
numerical techniques accounting for noisy photometry, irregular temporal
sampling and seasonal monitoring gaps. These techniques must also
account for the extrinsic variability in the quasar images, related to
microlensing effects, to avoid systematic error on the time-delay
measurements. Different techniques have been devised in the literature to carry
out this task, and the COSMOGRAIL collaboration has implemented its own
approach and several algorithms \citep[see][also for a summary of
extrinsic variability causes]{Tewes2013a}. These techniques are publicly
available as a python package named {\tt PyCS}\footnote{{\tt PyCS} can
be obtained from \url{http://www.cosmograil.org}}. They have been tested
using realistic numerical simulations, and have been confronted with the
data provided to the lensing community by the first Time-Delay
Challenge \citep[TDC1; see][]{Dobler2015, Liao2015}. An in-depth
analysis of their performance proved them to be both precise and
accurate \citep[][]{Bonvin2016} under various observational conditions,
and in particular for light curves mimicking the COSMOGRAIL data. Among
the three point-estimation algorithms provided in the {\tt PyCS}
toolbox, we consider for the present work two algorithms based on very
different principles:

\begin{enumerate}

\item {\bf The free-knot spline technique} models the light curves as a
sum of intrinsic variations of the quasar, common to the four light
curves, plus some extrinsic variability different in each of the four
light curves. The algorithm simultaneously fits one continuous curve for
the intrinsic variations, four less-flexible curves for the extrinsic
variations, and time shifts between the four light curves. All curves
are represented as \emph{free-knot} splines
\citep[see e.g.][]{Molinari2004}, for which the knot
locations are optimised at the same time as the spline coefficients and
the time shifts.

\item {\bf The regression difference technique} minimises the
variability of the difference between Gaussian-process regressions
performed on each light curve. This method has no explicitly
parametrised model for extrinsic variability. Instead, it yields
time-delay estimates which minimize apparent extrinsic variability on
time scales comparable to that of the precious intrinsic variability 
features. We see the contrasting approaches of this technique and the free-knot 
splines as valuable to detect potential method-related biases, and will use the 
regression difference technique 
as a cross-check of our results in this paper.

\end{enumerate}

The third original {\tt PyCS} estimator, a dispersion technique that was
inspired by \citet{Pelt1996} and used in the previous analysis of
\hequad\ \citep{Courbin2011} has proven to be less accurate in several investigations of
simulated data \citep[see][]{Eulaers2013, Rathnakumar2013, Tewes2013a,
Tewes2013b}. For this reason, we do not consider it in the present work.

We stress that the uncertainty estimation for the time delays is at
least as important as the above point estimators. It is carried out
within {\tt PyCS} by assessing the point-estimation performance on
synthetic light curves. This approach attempts to capture significantly
more than the formal uncertainty which could be derived from the
photometric error bars, if one would assume that for instance the spline
model described above is a sufficient description of the data.

\begin{figure*}
  \centering
    \includegraphics[width=0.97\textwidth]
{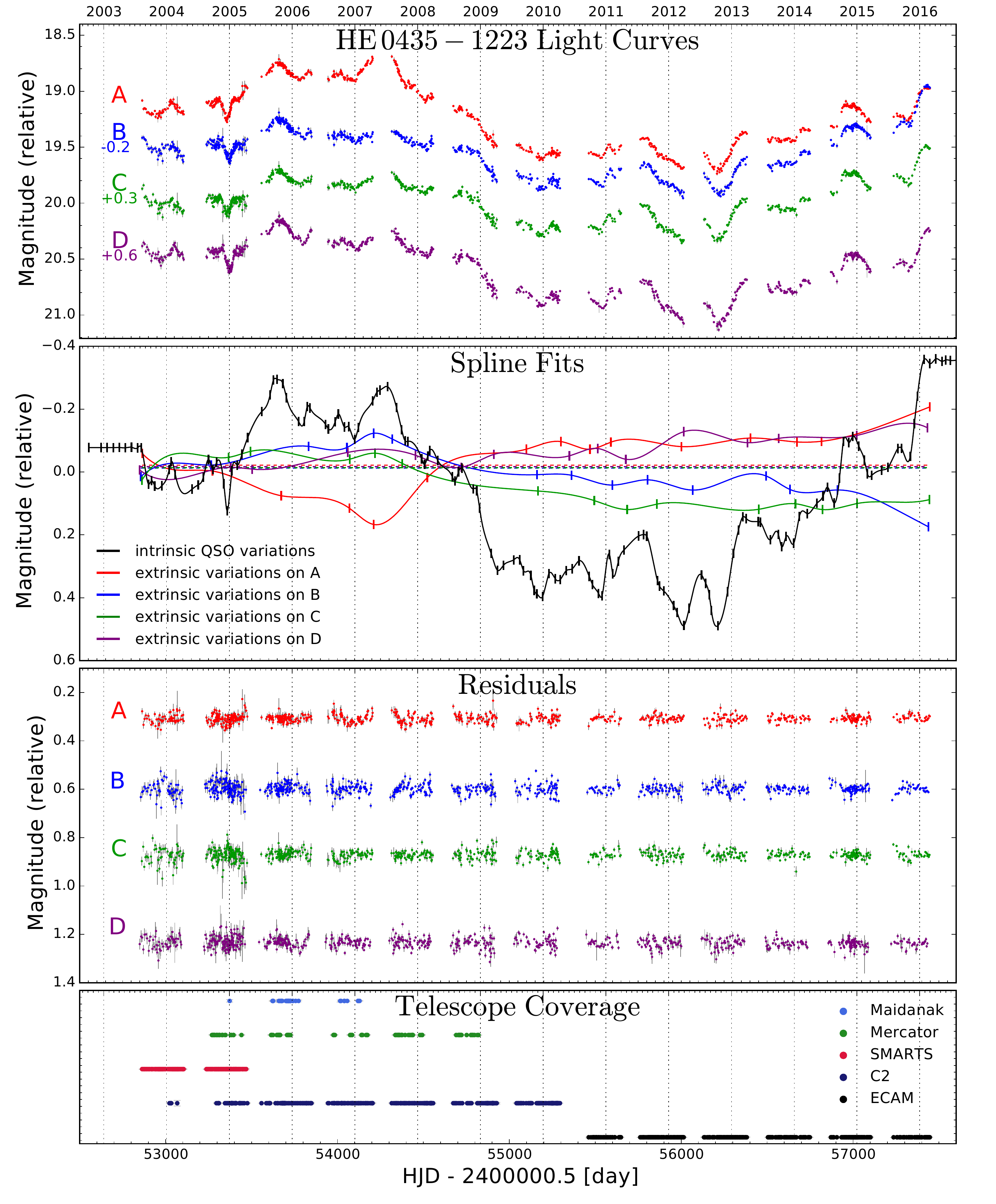}
    \caption{{\it From top to bottom}: Light curves for the four lensed
    images of the quasar \hequad. The relative shifts in magnitude are
    chosen to ease visualization, and do not influence the time-delay
    measurements. The second panel shows a model of the intrinsic
    variations of the quasar (black) and the 4 curves for the extrinsic
    variations in each quasar image using the free-knot spline
    technique (color code). The vertical ticks indicate the position of
    the spline knots. The residuals of the fits for each light curve is
    shown in the next panel. Finally, the bottom panel displays the
    journal of the observations for \hequad\ for the 5 telescopes or
    cameras used to gather the data over 13 years (see column
    ``$\#$obs'' of Table~\ref{tab:monitoring}), where each point
    represents one monitoring night. The light curves will be made
publicly available on the CDS and COSMOGRAIL websites once the paper is
accepted for publication.}
    \label{fig:he0435_lightcurve}
\end{figure*}

\begin{figure*}
  \centering
    \includegraphics[width=0.90\textwidth]{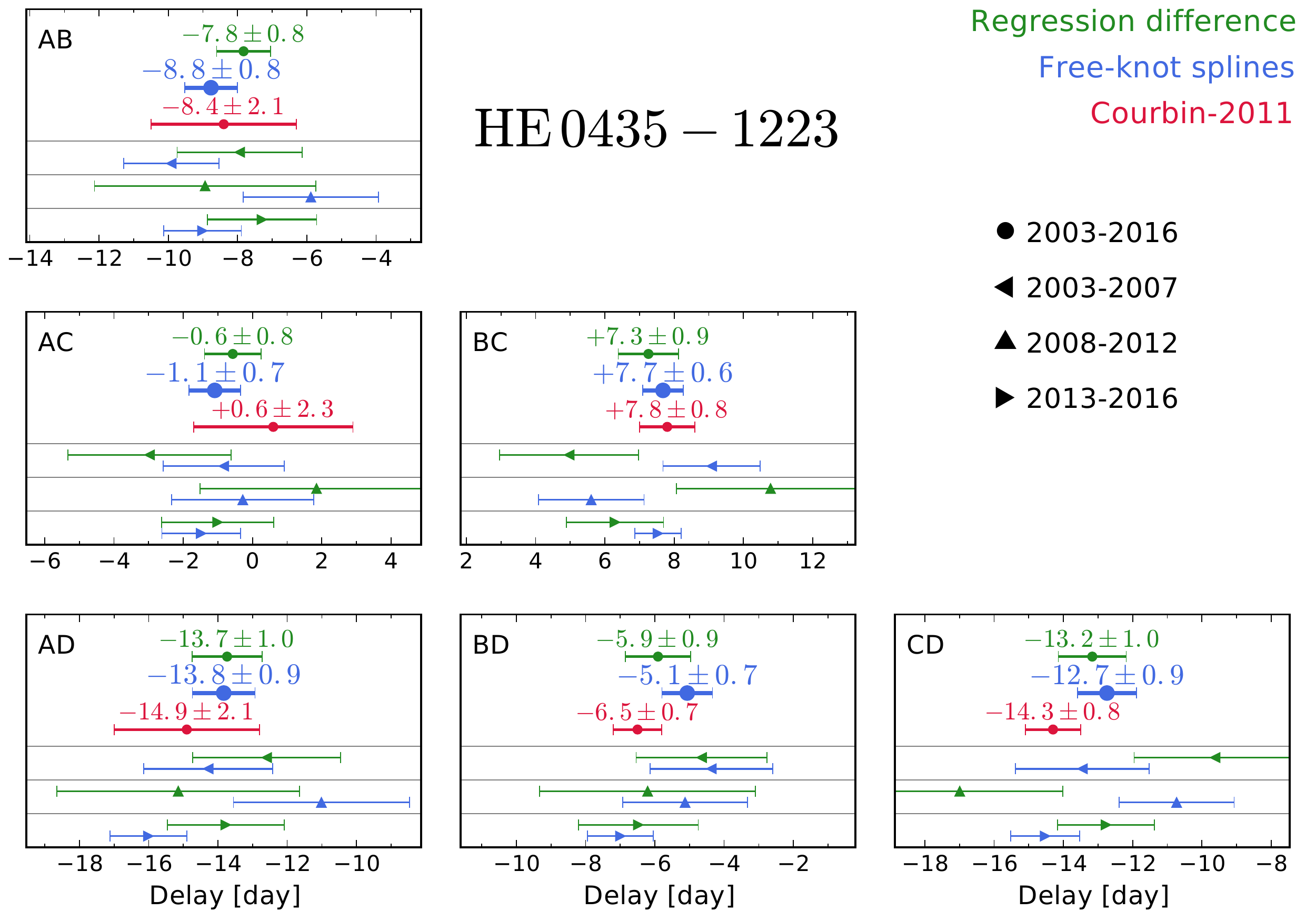}
    \caption{Time delays for the 6 pairs of quasar images, as indicated
    in top left corner of each panel. In each panel, we show the time
    delay measurement along with the 1$\sigma$ error bar using our two
    best curve-shifting techniques, and we compare with the measurement
    of \citet{Courbin2011}. We also show the result of measurements carried out with the
    free-knot spline technique and regression difference technique when
splitting the data in 3 continuous
    chunks of 4 or 5 years each. All cosmological results in this work
    use the time delay measurements from the free-knot splines (larger
    blue symbols on the figure).
%The scripts used to produce
%the presented time delays will be made publicly available on the
%COSMOGRAIL website once the paper is accepted for publication.
}
\label{fig:he0435_delays}
\end{figure*}

\subsection{Application to the data}

To apply the free-knot spline and regression difference techniques
provided by {\tt PyCS} to our data, we closely follow the procedure
described in \citet{Tewes2013a}, and summarized in the following.\footnote{For the sake of reproducibility,
the complete {\tt python} code used to measure the delays is available
at \url{http://www.cosmograil.org}}  A key
ingredient of this approach is the careful generation of mock light
curves which are used to fine-tune and assess the precision and bias of
the point estimators. These simulations are fully synthetic, in the
sense that they are drawn from models with known time delays 
(hereafter \emph{true} time delays), and
yet they closely mimic the quasar variability signal and the extrinsic
variability from the observed data. The {\tt PyCS} free-knot spline
technique is used to create the generative models from which we draw
these simulations. For this, we start by fitting an intrinsic spline
with on average 10 knots per year and four extrinsic splines with 2
knots per year to the observations. These average knot densities are
sufficiently high to fit all unambiguous patterns observed in the data,
while still resulting in a negligible intrinsic variance, i.e., avoiding
significant degeneracies between the time-delay estimates and the spline
models. Such a free-knot spline fit is illustrated in the second panel
of Figure~\ref{fig:he0435_lightcurve}.
The extrinsic variability splines presented here, 
when subtracted to each other pair-wise, are compatible with the data 
presented in \citet{Blackburne2014}. However, we also show in our 
robustness checks (see point (ii) of Section \ref{sec:robustness}) that 
variations in the modeling of the extrinsic variability do not influence 
much the time-delay measurements.

Before drawing the synthetic mock
curves by sampling from this model fit, the smooth extrinsic splines are
locally augmented with fast correlated noise. This noise follows a
power-law spectrum which is iteratively adjusted so that the scatter in
the mock curves has similar statistical properties to the scatter
measured in the observed data. We then draw 1000 mock datasets, with
true time shifts uniformly distributed within $\pm 3$ days around our
best-fitting solution. This results in a range of $\pm 6$ days for the
true delays, largely covering all plausible situations for this lens
system. It is important to use simulations with various true time delays
to tune and/or verify the accuracy of the point estimators. Tests on
simulations with only a single true time delay would not probe bias and
precision reliably, especially as many time-delay estimators are prone
to responding unsteadily to the true delay.

The third panel of Figure~\ref{fig:he0435_lightcurve} shows the observed
residual light curves after subtraction of a free-knot spline fit, and
the bottom panel depicts the coverage by the different telescopes and
instruments. During the first 5 years of monitoring, 3 to 4 different
telescopes were used, with a mean residual dispersion of all data points
of $\sigma = 25$ mmag. During the last years (2011 to present) one
telescope was used, with a mean residual dispersion of $\sigma = 15$
mmag. Besides unmodeled microlensing effects, part of this scatter comes
from night-to-night and instrument-to-instrument calibration of the
data. Long-term monitoring programs of gravitational lenses are a matter
of balance between the gain in temporal sampling using multiple
telescopes, and the losses in photometric precision due to combining data
from different instruments. Future monitoring programs will need to
account for this trade-off (Courbin et al. 2016, in prep).

We run the free-knot spline fit and the regression difference technique
(with a Mat\'ern covariance function, an amplitude parameter of 2.0 mag,
a scale of 250 days, and a smoothness degree $\nu$ = 1.5) on the
observed light curves as well as on the mocks \citep[for details,
see][]{Tewes2013a}.
Figure~\ref{fig:he0435_delays} presents our time-delay estimates along
with their 1$\sigma$ uncertainties, and compares them to the previous
measurements by \citet{Courbin2011}, for which the dispersion technique
was used. The uncertainties are computed by summing
the maximum estimated bias and statistical uncertainty in quadrature.
The free-knot spline technique and regression difference technique are
in relatively good agreement with each other, with a maximum tension of
1.3$\sigma$. Recall that the measurements are not independent, and
therefore good agreement is to be expected. The two techniques also yield a
similar precision, with a 6.5$\%$ relative uncertainty on the longest
delay, i.e. $\Delta t_{\mathrm{AD}}$.

\subsection{Robustness checks}
\label{sec:robustness}

In order to test the robustness of our time-delay measurements, we
performed several simple checks:

\begin{enumerate}

\item{ We carried out several times the deconvolution of the ECAM data,
using PSF stars and/or normalization stars that differ from the ones
adopted in Figure~\ref{fig:he0435_deepfield}. We also changed the initial
parameters of the MCS deconvolution photometry. These include an
estimate for the light profile of the lens galaxy, the astrometry of the
quasar images and of lens galaxy and the flux of the quasar images at
each epoch. All these changes resulted in a slightly higher scatter in the
ECAM light curves data points, yet without significant impact on the
time delay measurements.}

\item{We varied the intrinsic and/or extrinsic variability model of the
free-knot spline technique by changing the number of knots
used. We used 8 to 12 knots per year for the intrinsic model, and
0.5 to 2 knots per year for the extrinsic model. Free-knot splines have
the advantage
over regular splines or polynomials that their ability to fit prominent
variability features is less sensitive to the total number of
parameters. Using a lower or higher number of knots did not
significantly affect the time-delay measurements. The residual light
curves (third panel of Figure~\ref{fig:he0435_lightcurve}) remain
statistically similar.}

\item{Taking advantage of the 13 years of monitoring, we split the light
curves into three parts: i) seasons 2003-2004 to 2006-2007, ii) seasons
2007-2008 to 2011-2012 and iii) seasons 2012-2013 to 2015-2016. We
measured the time delays independently on each of these subsections. The results are
presented in the bottom parts of each panel of
Figure~\ref{fig:he0435_delays}. We see that the measurements on these subsections are well
distributed around the delays measured on the full light curves.
Furthermore, a clear majority of the delays obtained on the subsections
cover, within the given $1\sigma$ error estimates, the results from the
full curves. To conclude, these robustness checks give no strong
evidence that the achieved time-delay uncertainties are significantly
underestimated and/or biased.
}

\end{enumerate}

\subsection{Time delays of \hequad}

We have shown that our two curve shifting techniques lead to comparable
time delays and error bars on the full light curves of \hequad, which is
reassuring. Still, one needs to define which time delay estimates to
propagate into the time-delay distance (\PaperIV) and
cosmological parameter inferences. We opt for using  the results from the free-knot
spline technique. This method
has been tested extensively on a broad range of simulated light curves
and proved to be both precise and accurate \citep{Bonvin2016}. In
addition, \citet{Sluse2014} showed with this same technique that a
flexible extrinsic variability model can prevent potential
time delay biases due to the delayed emission of the Broad Line Region
of the quasar with respect to the accretion disk.

%=========================================================
%			TIME DELAY DISTANCE
%=========================================================

\section{Time-delay distance}
\label{sec:modeling}

The time delays determined in Section~\ref{fig:he0435_delays}, combined
with a careful modeling of the lens galaxy mass distribution, can be
used to infer the
time-delay distance in the \hequad\ system. The lens modeling and 
time-delay
distance determination are addressed in detail in \PaperIV and
are only summarized here.

\subsection{Principles of the measurement}

The time delay $\Delta t_{ij}$ between two lensed images of the same
object can be
written as follows:
\begin{equation}\label{eq:tdd}
\Delta t_{ij} = \frac{D_{\Delta t}}{c} \left[
\frac{(\bm{\theta}_i - \bm{\beta})^2}{2} -
\psi(\bm{\theta}_i) - \frac{(\bm{\theta}_j -
\bm{\beta})^2}{2} +
\psi(\bm{\theta}_j)\right].
\end{equation}
where $\bm{\theta}_i$ and $\bm{\theta}_j$ are the
coordinates of the images $i$ and $j$ in the lens plane,
$\bm{\theta}$ is the position of the lensed images on the plane of
the
sky, $\bm{\beta}$ is the unlensed source position and
$\psi(\bm{\theta}_i)$ is the lens potential at position
$\bm{\theta}_i$. The time-delay distance $D_{\Delta t}$
is defined to be the following
combination of three angular diameter distances and the deflector (i.e. 
the lens) redshift $z_d$:
$D_{\Delta t} \equiv (1+z_d)
D_{\rm d} D_{\rm s} / D_{\rm ds}$.
Here, $D_{\rm d}$, $D_{\rm s}$ and $D_{\rm ds}$ are respectively the angular
distances between the observer and the deflector, the
observer and the source, and the deflector and the source.
The time-delay distance is, by
construction,
proportional to the inverse of the Hubble constant $H_0^{-1}$, and is
primarily sensitive to this of all cosmological parameters. A
posterior probability distribution for $D_{\Delta t}$ allows
us to infer a probability distribution for \hc, assuming a given
cosmology.

In the case of \hequad, there are multiple galaxies at
different redshifts close in projection to the strong lens system.
We explicitly include these galaxies in our multi-lens plane lens model
in \PaperIV, and in doing so introduce more angular diameter distances
into the problem. However, we can still form the posterior predictive
distribution for the ``effective'' time-delay distance defined above, 
and
it is the latter that we use to infer cosmological parameters.
All the remaining additional mass along the line-of-sight can also weakly
focus and defocus the light rays from the
source, an effect that needs to be corrected for. We model this external contribution
using an external convergence term $\kappa_{\rm{ext}}$ that modifies the
time-delay distance as follows:
\begin{equation}
D_{\Delta t} =\frac{D_{\Delta
t}^{\rm model} }{ 1-\kappa_{\rm{ext}}}.
\end{equation}
Here, $D_{\Delta t}^{\rm model}$ is the
effective time-delay distance predicted by the multi-plane model, and
$D_{\Delta t}$ is the corrected time-delay distance we seek. Given
probability density functions (PDFs)
for $P(D_{\Delta t}^{\rm model})$ and $\kappa_{\rm{ext}}$, we can 
compute the PDF for $D_{\Delta t}$.
In \PaperIV we derive a log-normal approximation for $P(D_{\Delta t})$,
and it is this that we use as a likelihood function $P(D_{\Delta t}|\boldsymbol{\theta},H)$ for cosmological parameters
$\boldsymbol{\theta}$ given a cosmological model~$H$.

In the rest of this section we provide a brief summary of each part of
the analysis just outlined, before proceeding to the cosmological
parameter inference in Section~\ref{sec:cosmography}.

\subsection{Determination of the external convergence}

We use two complementary approaches to quantify the impact of the mass
along the line-of-sight, both yielding consistent results.

\subsubsection{Spectroscopy of the field}

In \PaperII, we perform a spectroscopic identification of a
large fraction of the brightest galaxies\footnote{The completeness of
the spectroscopic identification depends on the distance to the lens and
limiting magnitude, see Figure~3 of \PaperII. For example, 60\%
(80\%) of the galaxies brighter than $i \sim 22\,$mag ($i \sim
21.5\,$mag) have a measured spectroscopic redshift within a radius of
3\arcmin\, (2\arcmin) of the lens.} located within a projected distance
of 3\arcmin\, of the lens. This catalog is complemented with
spectroscopic data from \citet{Momcheva2015} that augment redshift
measurements to projected distances of $\sim$15\arcmin\, from the 
lens. Based on those data, we show that, from the five galaxies located 
within 12 arcsec of the lens, the galaxy G1 ($z=0.782$), closest in 
projection, produces the largest perturbation of the gravitational 
potential, and hence needs to be explicitly included in the lens 
models. The other galaxies are found to produce significantly smaller 
perturbations. On the other hand, we search for galaxy groups and 
clusters that would
be massive enough to modify the structure of the lens potential, but
find none. On the lower mass end (i.e. groups with $\sigma \leq 500$\,
km s$^{-1}$), 9 group candidates are found in the vicinity of the lens.
We demonstrate that none of the groups discovered is massive
enough/close enough in projection to produce high order perturbation of
the gravitational lens potential \citep{McCully2014, McCully2016}. This
is also confirmed by a weak lensing analysis of the field of \hequad\
(Tihhonova et al., in prep.)

\subsubsection{Weighted galaxy number counts}
\label{sec:kappa_ext}

In \PaperIII, we calculate the probability distribution for the
external convergence using a weighted galaxy number counts technique
\citep{Greene2013}. We
conduct a wide-field, broad-band optical to mid-infrared photometric
survey of the field in order to separate galaxies from stars, determine
the spatial distribution of galaxies around \hequad, and
estimate photometric redshifts and stellar masses. We compare weighted
galaxy
number counts around the lens, given an aperture and flux limit, to
those through similar apertures and flux limits in CFHTLenS
\citep{Heymans2012}. We investigate weights that incorporate the projected distance
and redshift to the lens as well as the galaxy stellar masses. The
resulting number under/over-densities serve as constraints in selecting
similar fields from the Millennium Simulation, and their associated
$\kappa_{\rm{ext}}$ values, from the catalog of \citet{Hilbert2009}. We
find that the resulting distribution of $\kappa_{\rm{ext}}$ is
consistent with the typical mean density value (i.e.
$\kappa_{\rm{ext}}$=0) and is robust to choices of weights, apertures,
flux limits and cosmology,
up to an impact of ~0.5\% on the
time-delay distance.

\subsection{Mass modeling}
\label{sec:mass_modeling}

In \PaperIV, we perform our lens modeling using \GLEE, a
software package developed by A.~Halkola and S.~H.~Suyu \citep{Suyu2010a,
Suyu2012}. Our fiducial mass model for the lens galaxy
is a singular power-law elliptical mass distribution with external
shear. We explicitly include the closest line-of-sight perturbing
galaxy in the lens model (G1; see Figure~3 of \PaperIV), using the full multiplane lens equation to
account for its effects. We also include in an extended modeling four
other
nearby perturbing galaxies to check their impact. Because the perturbers
are at different redshifts, there is no single time-delay distance that
can be clearly defined. Instead, we vary \hc directly in our models and
then use this distribution to calculate an effective time-delay
distance, where the angular diameter distances $D_{\rm d}$ and $D_{\rm
s}$ are calculated using the redshift of the main deflector, $z_{\rm d} =
0.454$.  We assume a fiducial cosmology, $\Omega_{m} = 0.3$, $\Omega_{\Lambda} = 0.7$,
and $w = -1$ in this modeling procedure, but we find that allowing
these cosmological
parameters to vary has a negligible ($< 1\%$) effect on the resulting
effective time-delay distance distribution.

The mass sheet degeneracy---the invariance to the lensed images under addition of a
uniform mass sheet to our mass model combined with a rescaling of the
source plane coordinates---can affect the inferred time delays, and
may limit the effectiveness of
time delays in constraining cosmology \citep[e.g.][]{Schneider2013,
Schneider2014}. We have shown in previous work that including the
central velocity dispersion of the
main galaxy in the lens modeling minimizes the effect of the mass sheet
degeneracy \citep[see Figure~4 of][]{Suyu2014}. In the case of \hequad\,
we measure $\sigma=222\pm15\, \rm{km\, s^{-1}}$ using Keck~I
spectroscopy. We also show that mass models that go beyond the elliptically-symmetric
power-law profile, and that are better physically justified, fit
our data equally well yet lead to the same cosmological inference. As in
\citet[][]{Suyu2014}, the \PaperIV tests both power-law and a
composite model with a baryonic component and a NFW dark matter halo. We
also note that the completely independent models of \citet{Birrer2016}
confirm the findings of \citet{Suyu2014}.

We model the images of the lensed source simultaneously in three \hst\
bands: ACS/F555W, ACS/814W, and WFC3/F160W. The lensed quasar
images are modeled as point sources convolved with the PSF. The extended,
unlensed image of the
host galaxy of the quasar is modeled separately on a pixel grid with
curvature regularization \citep[see e.g.][]{Suyu2006}. Our constraints
on the model include the positions
of the quasar images, the measured time delays, and the surface
brightness pixels in each of the three bands.  Model parameters of the
lens are explored through Markov Chain Monte Carlo (MCMC)
sampling, while the Gaussian posterior PDF for the source pixel values
is characterized using standard linear algebra techniques~\citep[e.g.][]{Suyu2006}.

During our modeling procedure, we iteratively update the PSFs using the
lensed AGN images themselves in a manner similar to \citet{Chen2016}, and
we use these corrected PSFs in our final models (for more details, see
Suyu et al. in preparation). 
We conduct multiple 
robustness tests to account 
for various systematic
uncertainties in the modeling. We vary our choice of modeling regions 
and use 
various weights for each pixel. We use various assumptions for the
light profiles fits of the lens galaxy and we model the lens 
using alternative mass models, comparing the use of power-law profiles 
and 
chameleon profiles. We also explicitly include the five 
nearest perturbing galaxies into our modeling. All the models are 
given a similar weight, reflecting the possible choices available 
through the analysis, and are combined together to yield a single
posterior PDF
for $D_{\Delta t}$. Fig.\,9 of 
\PaperIV presents the individual and combined posterior distributions, 
highlighting on one hand the relatively good agreement between the 
models, and on the other hand the need to consider a sufficiently 
flexible model to fully take into account as many sources of 
systematics as 
possible.

\subsection{Blinding methodology and unblinded results}
\label{sec:blindness}

A key element of our analysis is that it is carried out blindly with
respect to the inference of cosmological parameters. This blindness is
crucial in order to avoid unconscious confirmation bias. In practice, blindness is
built into our measurement in the following manner. All the individual
measurements and modeling efforts in \PaperIV are carried out without any knowledge
of the effects of specific choices on the resulting cosmology. In some
cases, this blindness is trivial to achieve: for example the measurement
of velocity dispersion was carried out and finalized independently from
the cosmological inference, and the connection between the two is
significantly complex and indirect that the person carrying out the
velocity dispersion measurement effectively had no way to determine how
that could affect cosmological parameters. In other cases, building on
the procedure established by our previous analysis of \rxjlens\
\citep{Suyu2013}, blindness was achieved by only using plotting codes that
offset every posterior probability distribution for time-delay distance 
and
cosmological parameters by a constant (such as the median value of each
marginal distribution), and thus never revealing the actual measurements to the
investigators
until the time of unblinding (see discussion in \PaperIV).

All of our
analysis and visualization tools were developed and tested using
simulated quantities. No modifications were allowed after the official
unblinding, making the unblinding step irreversible. The official
unblinding was originally scheduled for June 2 2016 during a
teleconference open to all the co-authors. Additional tests were
suggested during this meeting. As a result, the
analysis was kept blind for another two weeks and the final unblinding
happened during a teleconference starting at 6AM UT on June 16 2016
and was audio recorded by LVEK without others knowing 
until the end of 
the teleconference. The
results presented in the next section are the combination of the blind
measurements obtained for \hequad\ and \rxjlens\footnote{The
time-delay distance measurement of \rxjlens\ from \citet{Suyu2014} that
includes a composite model for the lens was not blind, whereas the
first measurement of this same lens from \citet{Suyu2013} was done
blindly.}, and the not-blind
measurements obtained by our team for the first system \blens.

%=========================================================
%			COSMOGRAPHY
%=========================================================

\section{Joint Cosmography Analysis}
\label{sec:cosmography}

CMB experiments provide a model-dependent value of the Hubble constant,
\hc, which appear to be in some tension with methods based on standard rulers and
standard candles. In a flat \LCDM universe, the significance of the
tension between the most recent values from Planck
\citep{Planck2015} and the direct measurement from Cepheids and Type Ia Supernovae
\citep{Riess2016} is 3.3$\sigma$. Either this tension is due, at least
in part, to systematics in the measurements
\citep[as suggested by e.g.][]{Efstathiou2014}, or it is caused by new physics beyond the
predictions of flat \LCDM. Several authors discuss the possibility of
relaxing the usual assumptions about cosmological parameters as a way to
reduce the tension \citep[e.g.][]{Salvatelli2013, Heavens2014,
DiValentino2016}. Possible assumptions include, for example, that we live in a
non-flat universe ($\Ok \neq 0$), that the dark energy equation of state
is not a cosmological constant ($w \neq -1 $), that the sum of the
neutrino masses is larger than
predicted by the standard hierarchy scenario (\mnu~$> 0.06$~eV), and/or
that the effective number of relativistic neutrino species may differ
from its assumed value in the standard model (\nnu\ $ \neq 3.046$).
Given
the above, it is important to consider a range of plausible extended
cosmological models when investigating the information that can be gained
from any specific cosmological probe \citep[see e.g.][]{Collett2014,
Giusarma2016}.

In this context, we present below our inference of the cosmological
parameters obtained using the time-delay distance measurements of the
strongly lensed quasars \blens, \wfilens\ and \hequad. After making
sure that their individual results are consistent
with each other, we present our cosmological inference using all three
systems jointly, referred as ``TDSL'' for ``Time Delay Strong Lensing.'' We
then combine TDSL with the WMAP Data Release 9
\citep[][hereafter ``WMAP'']{Bennett2013, Hinshaw2013} and with the Planck
2015 Data Release\footnote{
We use the Planck chains designated by ``plikHM\_TT\_lowTEB''
that uses the baseline high-L Planck power spectra and low-L
temperature and LFI polarization.}
\citep[][hereafter ``Planck'']{Planck2015}. When available, we also use the
combination of Planck data with Planck measurements of CMB weak-lensing
\citep[][hereafter ``CMBL'']{PlanckLensing2015}, with Baryon Acoustic
Oscillations surveys at various redshifts  \citep[][hereafter
BAO]{Percival2010, Beutler2011, Blake2011, Anderson2012,
Padmanabhan2012} and with the data of the Joint Lightcurve Analysis of
Supernovae \citep[][hereafter ``JLA'']{Betoule2013}. The latter two datasets are
described in detail in Section~5.2 of \citet{Planck2013} 
and Section~5.3
of \citet{Planck2015} respectively. Note that when possible, we do not
combine
the cosmological probes other than TDSL ourselves, but instead we use the
combined results published and provided by the Planck
team\footnote{\url{http://pla.esac.esa.int/pla/\#cosmology}}.

We follow the importance sampling approach suggested by \citet{Lewis2002} and
employed by \citet{Suyu2010b} and \citet{Suyu2013}, re-weighting the 
WMAP and Planck posterior samples with the TDSL likelihoods
from the analyses of \blens\
\citep{Suyu2010b},
\rxjlens\ \citep{Suyu2014} and \hequad\ (\PaperIV), for the set
cosmological models described in
Table~\ref{tab:cosmologies}.

\begin{table}
  \caption{Description of the cosmological models considered in this
  work. WMAP refers to the constraints given in the WMAP Data Release 9.
  Planck refers either to the constraints from Planck 2015 Data Release
  alone, or combined with CMBL, BAO and/or JLA. See
  Section~\ref{sec:cosmography} for details.}
  \centering
 \begin{tabular}{c | c}
 \hline
  Model name & Description\\
  \hline
  \UHO & \specialcell[t]{Flat \LCDM cosmology \\$\Om=1-\OL=0.32$ \\
\hc uniform in [0, 150]}\\
  \\%\hline
  \ULCDM & \specialcell[t]{Flat \LCDM cosmology \\$\Om=1-\OL$ \\ \hc
uniform in [0, 150]\\ $\Om$ uniform in [0, 1]}\\
  \\%\hline
  \UwCDM &  \specialcell[t]{Flat \wCDM cosmology \\ \hc uniform in [0,
150] \\ $\Ode$ uniform in [0, 1] \\ $w$ uniform in [$-$2.5, 0.5]}\\
  \\%\hline
  \UoLCDM &  \specialcell[t]{Non-flat \LCDM cosmology \\$\Om=1-\OL-\Ok
> 0$ \\ \hc uniform in [0, 150] \\ $\OL$ uniform in [0, 1] \\ $\Ok$
uniform in [$-$0.5, 0.5]}\\
  \\%\hline
  \oLCDM & \specialcell[t]{Non-flat \LCDM cosmology \\WMAP/Planck for
\{\hc, $\OL$, $\Om$\} \\ $\Ok=1-\OL-\Om$}\\
  \\%\hline
  \nnuLCDM & \specialcell[t]{Flat \LCDM cosmology \\WMAP/Planck for
\{\hc, $\OL$, \nnu\}}\\
  \\%\hline
  \mnuLCDM & \specialcell[t]{Flat \LCDM cosmology \\WMAP/Planck for
\{\hc, $\OL$, \mnu\}}\\
  \\%\hline
  \wCDM & \specialcell[t]{Flat \wCDM cosmology \\Planck for \{\hc, $w$,
$\Ode$ \}}\\
  \\%\hline
  \nnumnuLCDM & \specialcell[t]{Flat \LCDM cosmology \\Planck for
\{\hc, $\OL$, \mnu, \nnu \}}\\
  \\%\hline
  \owCDM & \specialcell[t]{Open \LCDM cosmology \\Planck for \{\hc,
$\Ode$, $\Ok$, $w$\}}\\
  \hline
 \end{tabular}

 \label{tab:cosmologies}

\end{table}

\begin{table*}
 \caption{Parameters of the three strong lenses used in our analysis.
 $\mu_D, \sigma_D$ and $\lambda_D$ are related to the analytical fit of
 the time-delay distance probability function (see 
Eq.~\ref{eq:pdt}).}
  \begin{tabular}{l c c c c c c}
  \hline
  Name & Reference & $z_d$ & $z_s$ & $\mu_{D}$ & $\sigma_{D}$ &
$\lambda_{D}$  \\
  \hline
  \blens\ & \citet{Suyu2010b} & 0.6304 & 1.394 & 7.0531 &  0.22824
& 4000.0\\
  \rxjlens\ & \citet{Suyu2014} & 0.295 & 0.654 & 6.4682 &  0.20560 &
1388.8\\
  \hequad\ & \PaperIV & 0.4546 & 1.693 & 7.5793 &
0.10312 & 653.9 \\
  \hline
  \end{tabular}
  \label{tab:lensparams}
\end{table*}

We consider both i) ``uniform'' cosmologies, with
only a few variable cosmological parameters with uniform priors in order
to get constraints from TDSL alone, and ii) cosmologies extended beyond
\LCDM where we combine the TDSL likelihoods with other probes. When 
comparing two cosmological parameter inferences, we use the
following terminology. When two results differ by less than 1$\sigma$ we
consider that they are ``consistent;'' in the 1-2$\sigma$ range they are in
``mild tension;'' in the 2-3$\sigma$ range they are in ``tension;'' above
$3\sigma$, they are in ``significant tension.'' If a cosmological
parameter inference follows a non-Gaussian distribution, we use 
``1$\sigma$''
to refer to the width of the distribution between its 50th and 16th
percentiles if the comparison is made towards a lower value, and between
its 50th and 84th percentiles if the comparison is made towards a higher
value. In a comparison, the $\sigma$ values always refer to those belonging to the
distribution that includes TDSL.

\subsection{Cosmological inference from Strong Lensing alone}\label{sec:ubgcosmo}

We first present our values for the cosmological parameters
that can be inferred from TDSL alone. We use the time-delay distance 
likelihoods
analytically expressed with a skewed log-normal distribution:
\begin{equation}\label{eq:pdt}
P(D_{\mathrm \Delta t}) =
\frac{1}{\sqrt{2\pi}(x-\lambda_{D})\sigma_{D}}
\mathrm{exp} \left[ -\frac{(\mathrm{ln}(x-\lambda_{D})-\mu_{D})^{2}}{2\sigma_{D}^{2}} \right],
\end{equation}
where $x=D_{\mathrm \Delta t}/(1\, \mathrm{Mpc})$. We recall
in Table~\ref{tab:lensparams} the lens and source redshifts of the three
strong lenses as well as the parameters $\mu_{D},\, \sigma_{D}$ and
$\lambda_{D}$ describing their respective
time-delay distance distributions.

\subsubsection{Combination of three lenses}

Before carrying out a joint analysis of our three lens systems, we first
perform a quantitative check that our three lenses can be combined without any loss
of consistency. For that purpose, we compare their time-delay distance
likelihood functions in the full cosmological parameter space, and
measure the degree to which they
overlap. Following \citet{Marshall2006} and \citet{Suyu2013}, we
compute the Bayes
factor $F$, or evidence ratio, in favor of a simultaneous fit of the
lenses using a common set of cosmological parameters. When comparing
three data
sets $\bf{d_1}, \bf{d_2}$ and $\bf{d_3}$, we can either assume the
hypothesis $\rm{H^{global}}$ that they can be represented using a
common global
set of cosmological parameters, or the hypothesis $\rm{H^{ind}}$ that
at least one data set is better represented using another independent
set of cosmological parameters. We stress that this latter model would 
make sense if
there was a systematic error present that led to a vector offset in
the inferred cosmological parameters. To parametrize this offset vector
with no additional information would take as many nuisance parameters as
there are dimensions in the cosmological parameter space; assigning uninformative
uniform prior PDFs to each of the offset components is equivalent to using
a complete set of independent cosmological parameters for the outlier 5
dataset.

The Bayes factor, that makes the $\rm{H^{global}}$ hypothesis
$F$ times more probable than $\rm{H^{ind}}$ can be
computed as follows:
\begin{equation}
 F=\frac{P(\bf{d_1}, \bf{d_2}, \bf{d_3} | \rm{H^{global}})}{P(\bf{d_1}|\rm{H^{ind}})P(\bf{d_2}|\rm{H^{ind}})P(\bf{d_3}|\rm{H^{ind}})}.
\end{equation}
\noindent A Bayes factor $F$ significantly larger than 1 indicates that
the considered data sets can be consistently combined. In the present
case,
considering three lenses with known time-delay distance likelihoods
$\bf{L_1}, \bf{L_2}$ and $\bf{L_3}$, the Bayes factor becomes:
\begin{equation}
 F_{\rm 1\cup2\cup3}=\frac{\langle \bf{L_1} \bf{L_2} \bf{L_3} \rangle}{\langle \bf{L_1} \rangle \langle \bf{L_2}\rangle \langle \bf{L_3}\rangle},
\end{equation}
\noindent where angle brackets denote averages over our ensembles of
prior samples. We can also compare the likelihoods pair by pair
(1-versus-1) as in Equation~27 of \citet[][]{Suyu2013} and then combine each
pair with the third likelihood (2-versus-1):
\begin{equation}
 F_{\rm 12\cup3}=\frac{\langle \bf{L_1} \bf{L_2} \bf{L_3}\rangle}{\langle \bf{L_1} \bf{L_2}\rangle \langle \bf{L_3}\rangle}.
\end{equation}
\noindent This last equation allows us to check that the lenses can also be well
combined pair-wise, and that none of them is inconsistent with the two
others considered together. We compute the Bayes Factors $F_{\rm 1\cup2\cup3}$
and all the possible 1-versus-1 and 2-versus-1 permutations in the
uniform cosmologies \UHO, \ULCDM, \UwCDM, and \UoLCDM. We find that all
the combinations are in good agreement, the only exception being for the
pair \blens~$\cup$~\rxjlens, which is only marginally consistent in the
\UoLCDM cosmology ($F_{\rm 1\cup2}$=1.1). Considering the likelihoods
individually, the three lenses are in excellent agreement, with a Bayes
Factor $F_{\rm 1\cup2\cup3}=21.3$ in \UHO, $F_{\rm 1\cup2\cup3}=14.2$ in \ULCDM,
$F_{\rm 1\cup2\cup3}=18.9$ in \UwCDM and $F_{\rm 1\cup2\cup3}=10.8$ in \UoLCDM.
We conclude that the time delay likelihoods of our three lenses can be
combined without any loss of consistency.

\subsubsection{Constraints in uniform cosmologies}

\begin{figure*}
  \centering
    \includegraphics[width=0.99\textwidth]{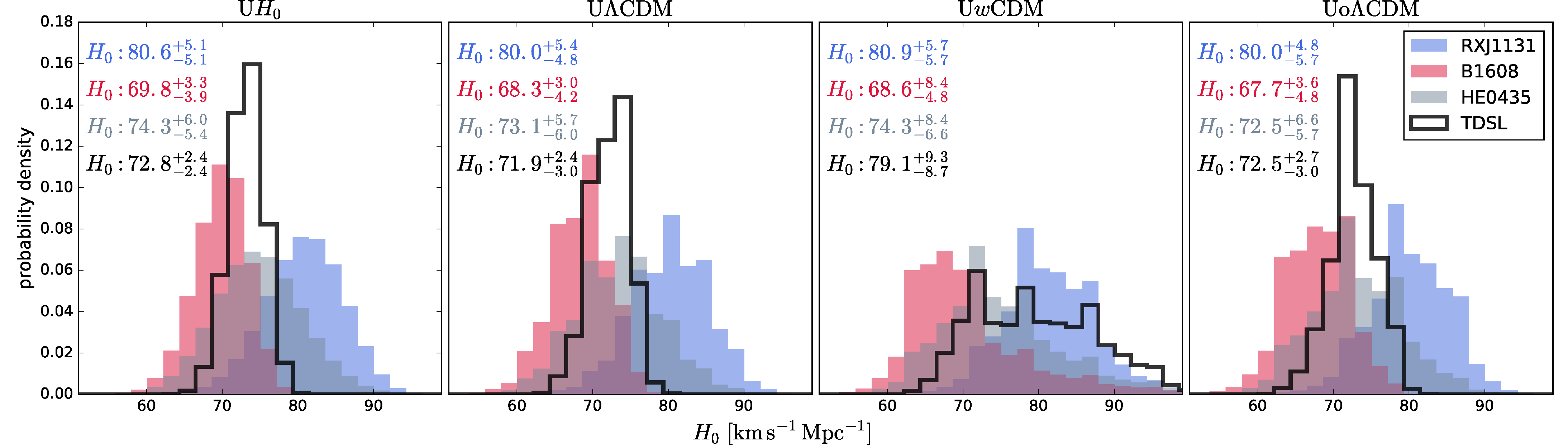}
    \caption{Marginalized posterior probability distributions for \hc
    in the \UHO, \ULCDM, \UwCDM and \UoLCDM cosmologies using the
    constraints from the three strong lenses \blens, \rxjlens\ and
    \hequad. The overlaid histograms present the distributions for each
    individual strong lens (ignoring the other two datasets), and the
    solid black line corresponds to the distribution resulting from the
    joint inference from all three datasets (TDSL). The quoted values of
    \hc in the top-left corner of each panel are the median, 16th and
    84th percentiles.} \label{fig:H0_ubgcosmo}
\end{figure*}

\begin{figure}
  \centering

\includegraphics[width=0.49\textwidth]{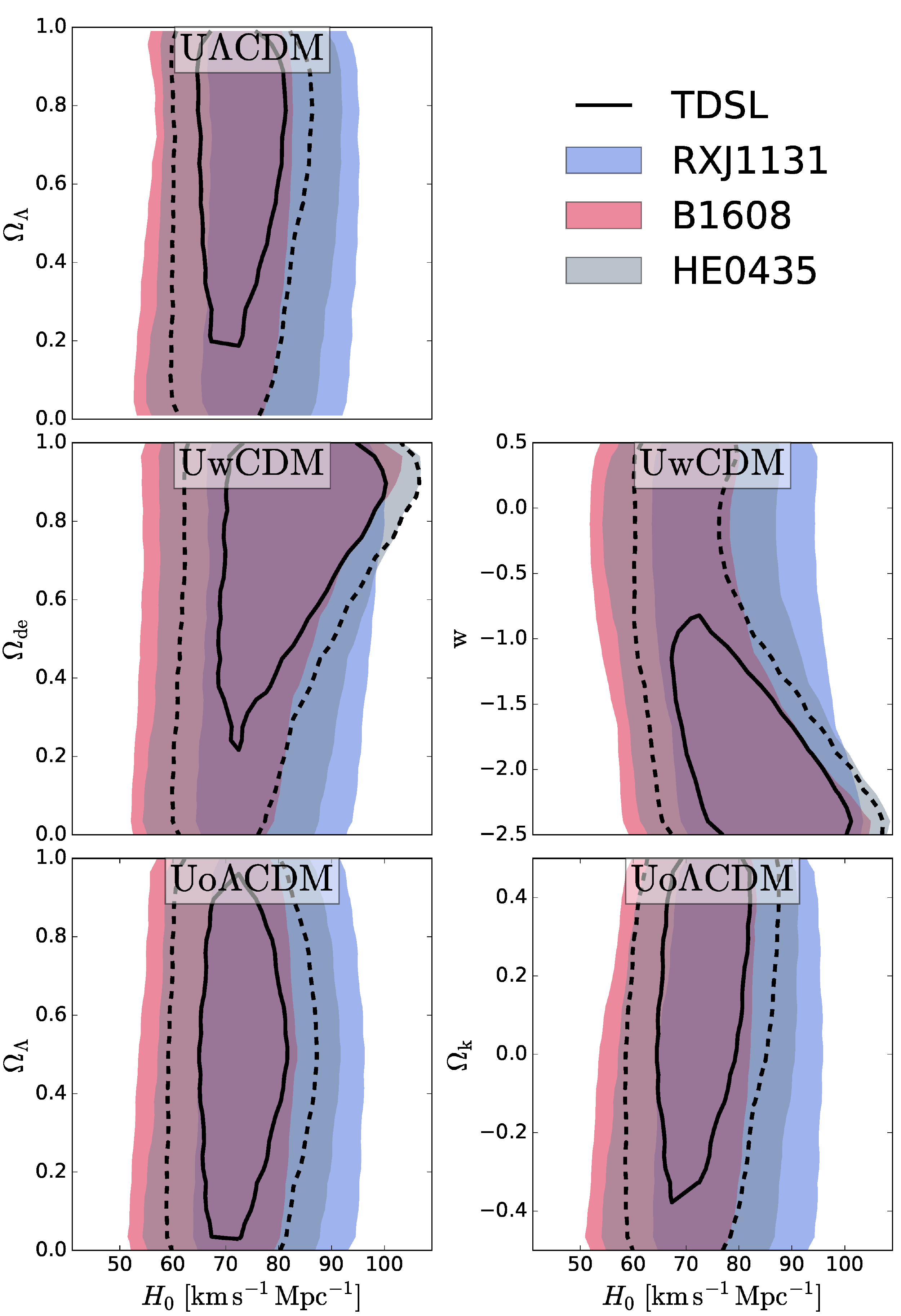}
    \caption{Comparison of the three strong lenses in the \ULCDM (top),
    \UwCDM (middle) and \UoLCDM (bottom) cosmologies. The colored
    overlays delimit the 95\% credible region for \blens, \rxjlens\ and
    \hequad. The solid and dashed black lines draw the contours of the
    68.3\% and 95\% credible  regions, respectively, for the combination
    of the three lenses. }
    \label{fig:H0-ode-w-ok_ubgcosmo}
\end{figure}

Figure~\ref{fig:H0_ubgcosmo} presents the marginalized posterior
PDF for \hc in the
cosmological models using uniform priors. Our baseline model, \UHO,
has a uniform prior on \hc in the range [0, 150] ${\rm km\,
s^{-1}\,Mpc^{-1}}$, a matter density fixed at $\Om=0.32$ from the most recent
Planck results \citep{Planck2015}, zero curvature $\Ok=0$ and consequently
a fixed cosmological constant. This model has only one free parameter.
From left to right in the figure, we
present this \UHO cosmology, and then three models that have two or three free parameters (\hc plus one or two others):
the \ULCDM cosmology where we allow $\Om$ to vary with uniform prior;
the
\UwCDM cosmology with a free $\Ode$ and a free time-independent dark energy equation of
state~$w$, both with uniform priors; and finally the \UoLCDM cosmology, that relaxes the
constraint on the curvature $\Ok$ and allows both this and $\OL$ to 
vary
with uniform priors. 
Table 
\ref{tab:cosmologies} summarizes the constraints and priors for these 
four 
models.
We quote in each panel the
corresponding median and 1$\sigma$ uncertainties of \hc. In the \UHO\
cosmology, combining the three lenses yields a value \hc${\rm
=72.8\pm2.4 \
km\, s^{-1}\,Mpc^{-1}}$, with 3.3$\%$ precision. When relaxing the
constraint on $\Om$ in \ULCDM (and thus being completely independent of
any other
measurement), we obtain \hc${\rm
=71.9_{-3.0}^{+2.4} \
km\, s^{-1}\,Mpc^{-1}}$, with 3.8$\%$ precision. These two estimates
are respectively
2.5$\sigma$ and 1.7$\sigma$ higher than the most recent Planck
measurement in a
flat-\LCDM universe \citep[\hc=$66.93\pm0.62\ {\rm km\,
s^{-1}\,Mpc^{-1}}$;][]{Planck2015}, in
excellent agreement with the most recent results using distance ladders
\citep[\hc=$73.24\pm1.74\ {\rm km\, s^{-1}\,Mpc^{-1}}$;][]{Riess2016},
and compatible with other
local
estimates \citep[see e.g.][]{Bonamente2006, Freedman2012, Sorce2012,
Gao2016}. Whether the tension between the local and cosmological
measurements of \hc comes from systematic errors or hints at new
physics beyond flat-\LCDM is currently a hot topic of discussion in the
community \citep[see e.g.][and references therein]{Efstathiou2014,
Planck2015, Rigault2015, Spergel2015, Addison2016, DiValentino2016,
Riess2016}.

Intriguingly, we note that the 
\hc 
values yielded by each system individually get larger for lower lens 
redshifts. So far, we cannot state if this comes from a simple 
statistical fluke, an unknown systematic error or hints towards an 
unaccounted physical property. The addition of two more lenses from the 
H0LiCOW sample will certainly help us in that regard.

Figure~\ref{fig:H0-ode-w-ok_ubgcosmo} presents the two-dimensional 95\%
credible regions of the
cosmological parameters in the \UwCDM, \UoLCDM, and \ULCDM cosmologies
for each lens individually and for their combination (TDSL). Time
Delay Strong
Lensing is primarily sensitive to \hc, and the tilt in the $H_{\rm
0}-\Omega_{\Lambda}$, $H_{\rm 0}-w$, and $H_{\rm 0}-\Ok$ planes
illustrates its weak sensitivity to the dark energy density, dark energy
equation of state and curvature density, respectively. TDSL alone
agrees both with a flat universe and a cosmological constant,
although on the latter the credible region extends deeply into the
phantom dark energy domain ($w<-1$). In the
\UwCDM cosmology the correlation between \hc and $w$ is more
prominent than in the other models, leading to a larger dispersion of
the \hc distribution in the corresponding panel of
Figure~\ref{fig:H0_ubgcosmo}.
This dispersion is more prominent for values of 
$w<-1$, since in such cases the variation of the density of 
dark energy becomes larger at low redshifts. Since our 
measurements are performed at the redshift of the lenses we observe, 
going back to redshift zero and \hc produces the degeneracy with $w$.

This highlights the fact that our
cosmological inferences in this cosmology are more sensitive to the
prior range we choose. Thus, the 
resulting parameter values must be
considered as indicative of a trend rather than as absolute
measurements. We summarize our
values for \hc, $\Ok$, $w$ and $\Om$ from TDSL alone in the top
section of Table~\ref{tab:cosmoparams}.

\subsection{Constraining cosmological models beyond \LCDM}

\begin{figure*}

  \begin{minipage}[l]{0.48\linewidth}
      \includegraphics[width=0.98\textwidth]{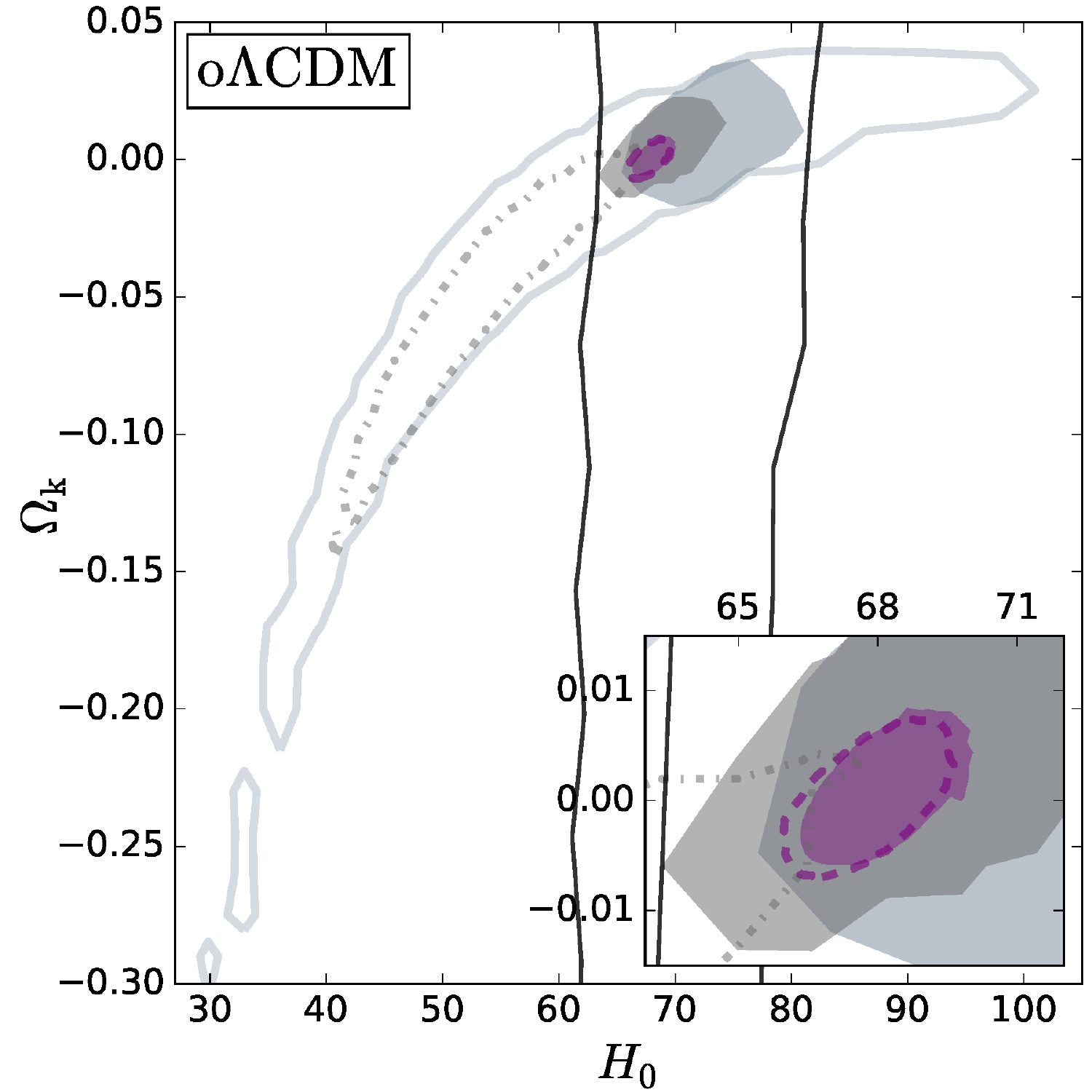}
  \end{minipage}
  \begin{minipage}[r]{0.48\linewidth}
      \includegraphics[width=0.98\textwidth]{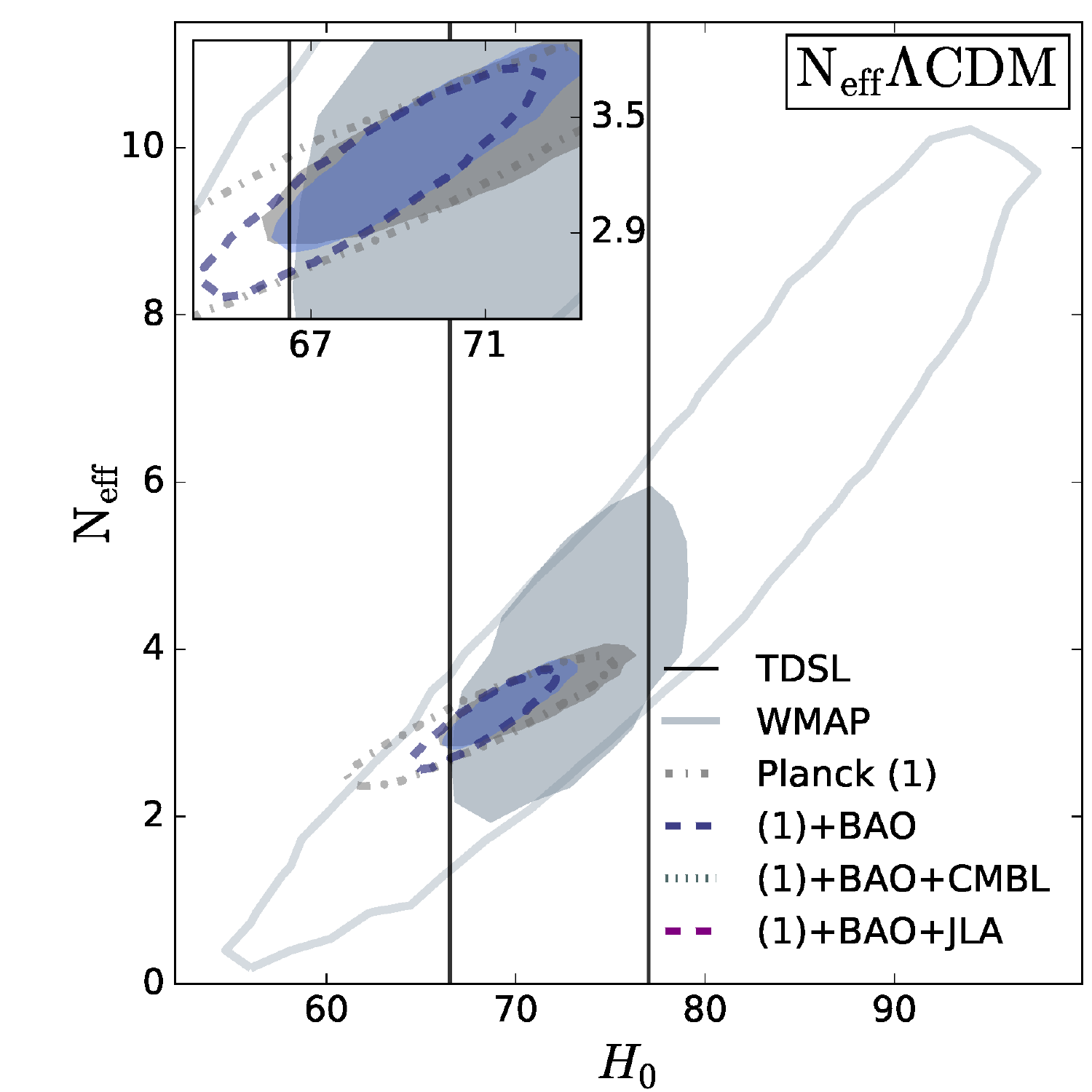}
  \end{minipage}

  \begin{minipage}[l]{0.48\linewidth}
      \includegraphics[width=0.98\textwidth]{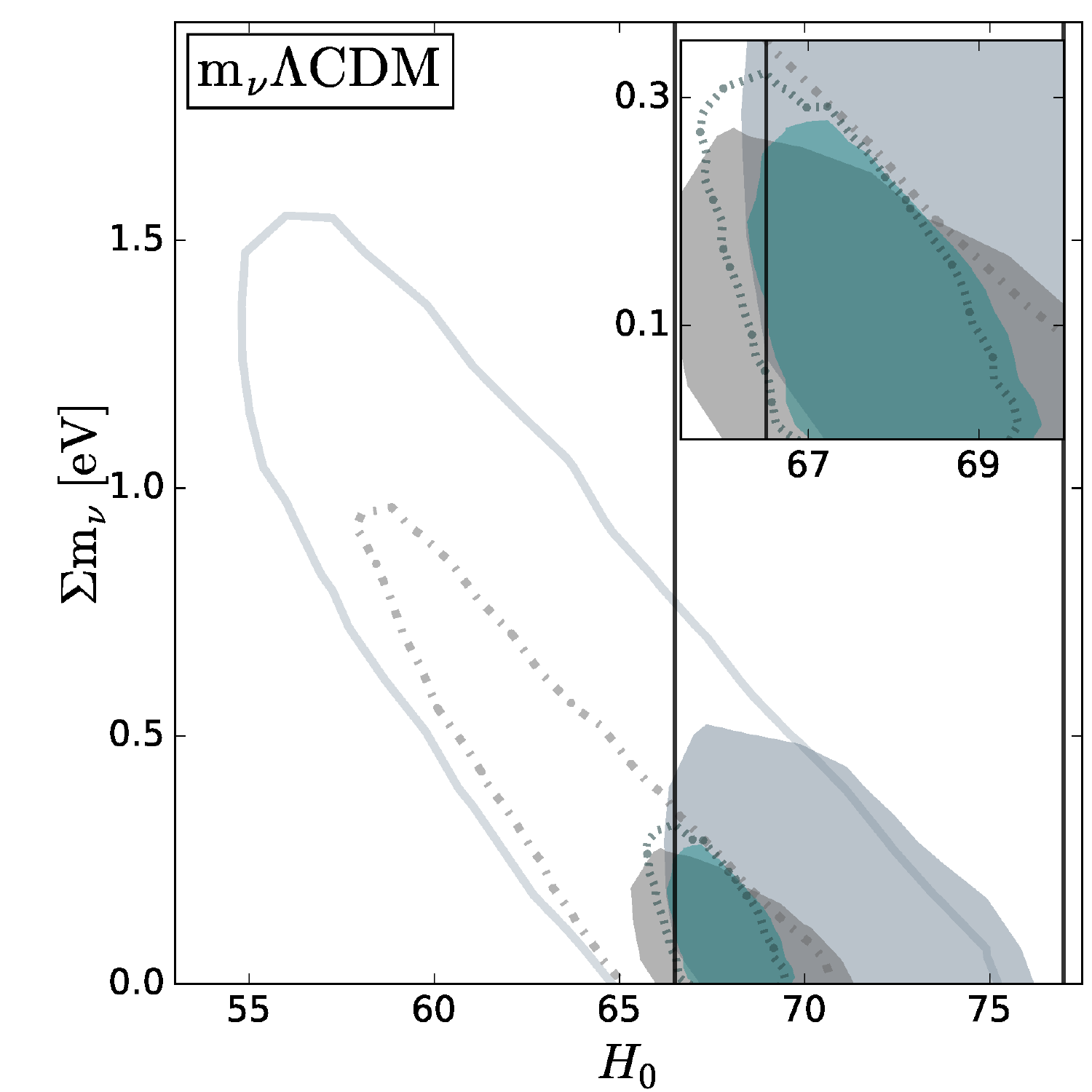}
  \end{minipage}
  \begin{minipage}[r]{0.48\linewidth}
      \includegraphics[width=0.98\textwidth]{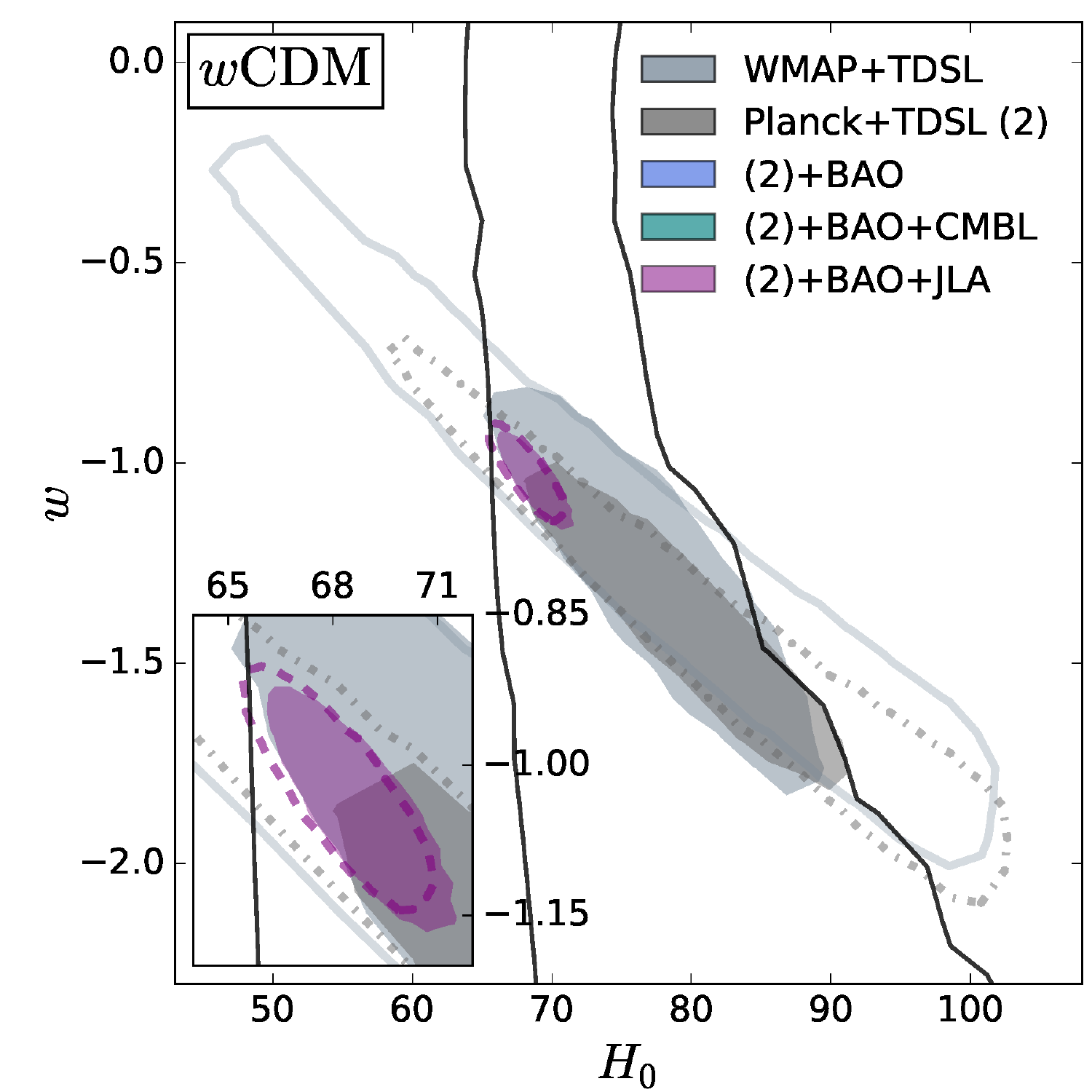}
  \end{minipage}

  \caption{Cosmological constraints in one parameter extensions to
\LCDM. We consider a non-flat universe with variable curvature $\Ok$
(top-left), a variable effective number of relativistic neutrino
  species \nnu (top-right), a variable total mass of
neutrino
  species \mnu (bottom-left, in eV) and a variable
time-invariant dark energy equation of state $w$ (bottom-right). The
filled regions and colored lines delimit the marginalized 95\% credible 
regions
(consistently smoothed due to the sparsity of the samples from the
available MCMC chains) with and without the constraints from TDSL
respectively. The different colors represent the constraints
from WMAP, Planck, Planck+CMBL, Planck+BAO, Planck+CMBL+BAO and
Planck+BAO+JLA. The
solid black lines delimit
the 95\% credible region for TDSL alone in the corresponding
uniform cosmology with no additional information.}

  \label{fig:cosmo_oneparamext}

\end{figure*}

\begin{figure*}

  \begin{minipage}{0.47\linewidth}
      
\centering\includegraphics[width=\linewidth]{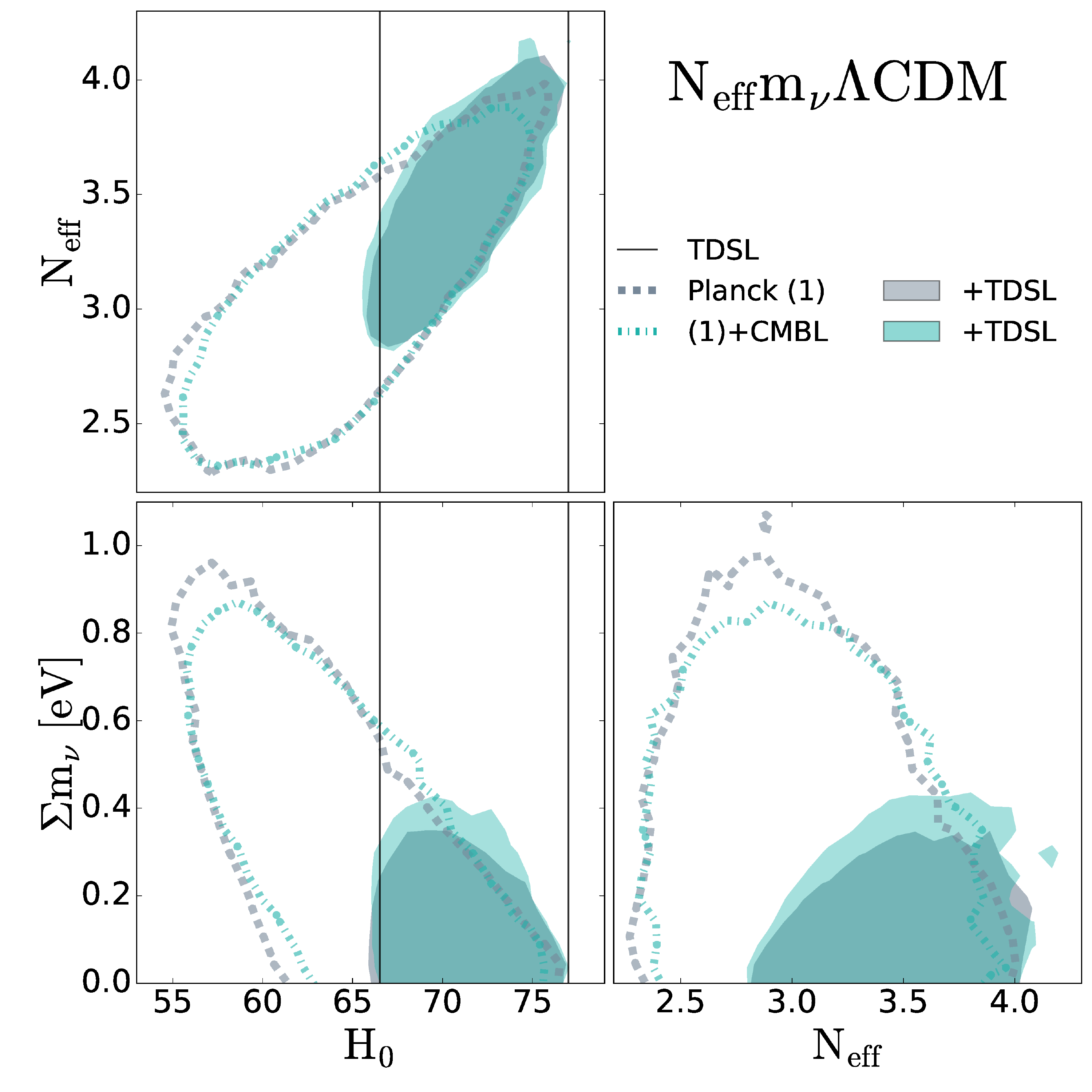}
  \end{minipage}\hskip 20pt
  \begin{minipage}{0.47\linewidth}
      \centering\includegraphics[width=\linewidth]{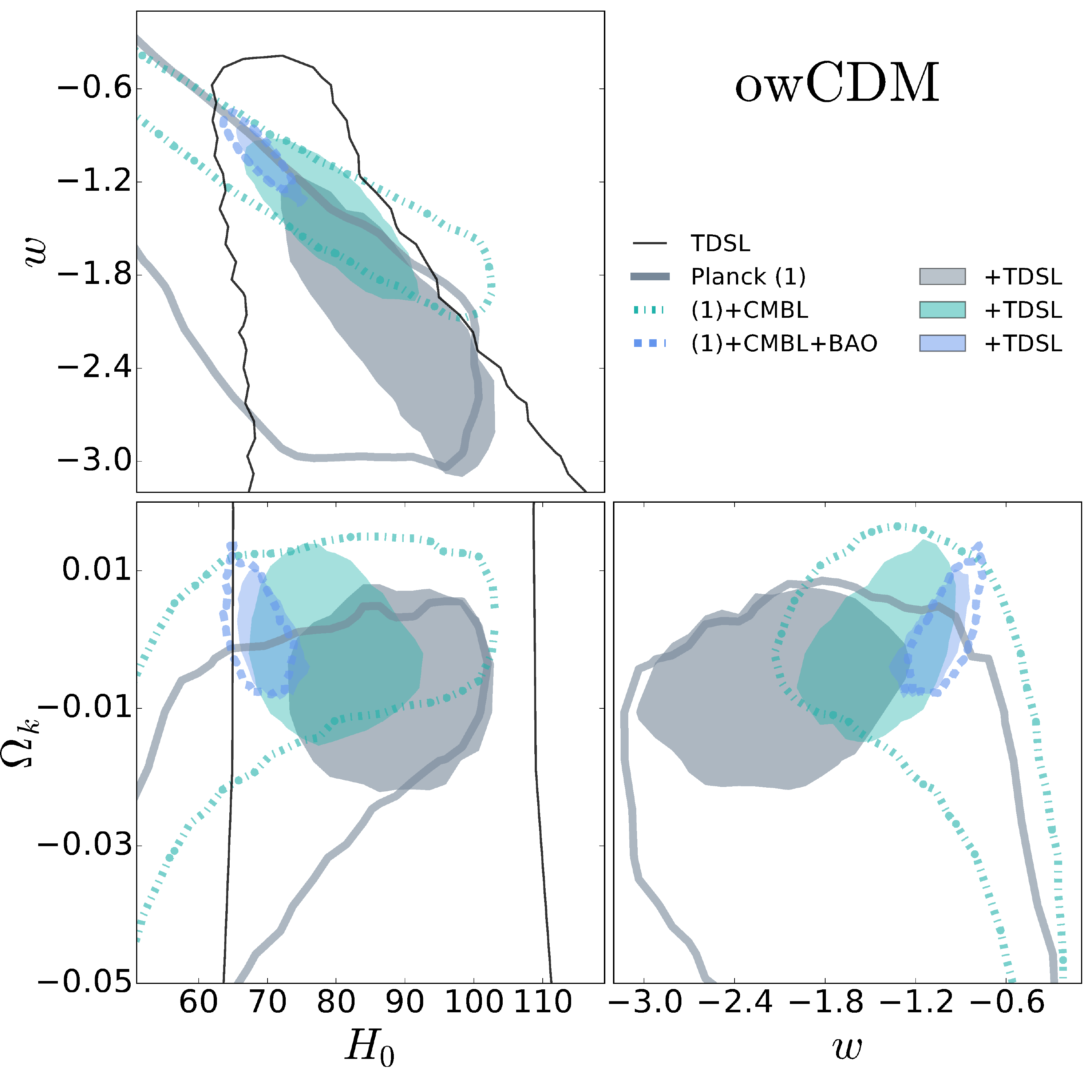}
  \end{minipage}

  \caption{Cosmological constraints in two-parameter extensions to
\LCDM. We consider a flat Universe with
  variable effective number of relativistic neutrino species \nnu and
  total mass of neutrinos \mnu (left), and an open universe with variable
  dark energy equation of state parameters $w$ (right). The colored
lines and filled areas are the same as in
Figure~\ref{fig:cosmo_oneparamext}, and show marginalized 95\% credible regions. The TDSL contours in
the \owCDM cosmology are computed using uniform priors on $\Ok$ [$-$0.5,
0.5]
and $w$ [$-$2.5, 0.5].}

  \label{fig:cosmo_twoparamext}

\end{figure*}

We now investigate how strong lensing can help constrain
cosmological models beyond standard flat \LCDM, when
combined with other cosmological probes. We demonstrated in
Section~\ref{sec:ubgcosmo} and Figure~\ref{fig:H0-ode-w-ok_ubgcosmo} that
TDSL
is only weakly dependent on the matter density, the dark energy
density, the dark energy equation of state and the
curvature. However, the cosmological parameter degeneracies for TDSL
are such that the combination of TDSL with other probes can rule out
large areas of parameter space. Following the motivations presented in
\citet{Planck2015} for extensions to the base \LCDM model, we present in
the following a selection of models where we combine TDSL with
the results from WMAP, Planck, Planck+BAO, Planck+BAO+CMBL and
Planck+BAO+JLA when available. Figures~\ref{fig:cosmo_oneparamext} and
\ref{fig:cosmo_twoparamext} present the results. Note that we have
smoothed the contours of the credible regions after importance
sampling with a gaussian filter due to the sparsity of 
the WMAP and Planck MCMC chains, checking
that the 95\% credible regions do indeed contain approximately
95\% of the importance weight.

\begin{table*}

 \caption{Summary of the cosmological parameters constraints for the
 models detailed in Table~\ref{tab:cosmologies}. \hc units are
 ${\rm km\ s^{-1}\ Mpc^{-1}}$, \mnu units are $\rm{eV}$. The quoted
 values are the median, 16th, and 84th percentiles, except
 for \mnu where we quote the 95\% upper bound of the probability
 distribution. The empty slots occur
 when no prior samples are provided by the Planck team.
}

 \begin{tabular}{l c c c c c c c c c c c c}
  \hline
  &\UHO & & \multicolumn{2}{c}{\ULCDM} & & \multicolumn{3}{c}{\UwCDM} &
& \multicolumn{3}{c}{\UoLCDM} \\
  & \hc & & \hc & $\Omega_{\Lambda}$ & & \hc & $\Ode$ & $w$
& & \hc & $\Omega_{\Lambda}$ & $\Ok$ \\
  \hline
TDSL & $72.8_{-2.4}^{+2.4}$ & & $71.9_{-3.0}^{+2.4}$ &
$0.62_{-0.35}^{+0.24}$ & & $79.1_{-8.7}^{+9.3}$ & $0.72_{-0.34}^{+0.19}$
& $-1.79_{-0.49}^{+0.94}$ & & $72.5_{-3.0}^{+2.7}$ &
$0.51_{-0.30}^{+0.28}$ & $0.1_{-0.3}^{+0.3}$ \\
  \hline
 \end{tabular}

 \vspace{0.5cm}

 \begin{tabular}{l c c c c c c c c}
  \hline
  &\multicolumn{4}{c}{\oLCDM} & & \multicolumn{3}{c}{\wCDM} \\
  & \hc & $\Om$ & $\Omega_{\Lambda}$ & $\Ok$ & & \hc & $\Ode$ & $w$ \\
  \hline
\vspace{0.1cm} TDSL+WMAP & $73.0_{-2.5}^{+2.3}$ & $0.25_{-0.02}^{+0.02}$ 
& $0.74_{-0.02}^{+0.02}$ & $0.005_{-0.005}^{+0.005}$ & & 
$76.5_{-3.9}^{+4.6}$ & $0.76_{-0.02}^{+0.02}$ & $-1.24_{-0.20}^{+0.16}$  
\\ 
\vspace{0.1cm} TDSL+Planck (1) & $69.2_{-2.2}^{+1.4}$ & 
$0.30_{-0.02}^{+0.02}$ & $0.70_{-0.01}^{+0.01}$ & 
$0.003_{-0.006}^{+0.004}$ & & $79.0_{-4.2}^{+4.4}$ & 
$0.77_{-0.03}^{+0.02}$ & $-1.38_{-0.16}^{+0.14}$  \\ 
\vspace{0.1cm} (1)+BAO & $68.0_{-0.7}^{+0.7}$ & $0.31_{-0.01}^{+0.01}$ & 
$0.69_{-0.01}^{+0.01}$ & $0.001_{-0.003}^{+0.003}$ & & 
$69.6_{-1.7}^{+1.8}$ & $0.70_{-0.01}^{+0.01}$ & $-1.08_{-0.08}^{+0.07}$  
\\ 
 (1)+BAO+JLA & $68.1_{-0.7}^{+0.7}$ & $0.31_{-0.01}^{+0.01}$ & 
$0.69_{-0.01}^{+0.01}$ & $0.001_{-0.003}^{+0.003}$ & & 
$68.8_{-1.0}^{+1.0}$ & $0.70_{-0.01}^{+0.01}$ & $-1.04_{-0.05}^{+0.05}$ 
\\ 
\hline
 \end{tabular}

 \vspace{0.5cm}

 \begin{tabular}{l c c c c c c c }
  \hline
  &\multicolumn{3}{c}{\nnuLCDM} & & \multicolumn{3}{c}{\mnuLCDM} \\
  & \hc &  $\Omega_{\Lambda}$ & \nnu & & \hc & $\Omega_{\Lambda}$ & \mnu (eV)
\\
  \hline
\vspace{0.1cm} TDSL+WMAP & $73.2_{-2.4}^{+2.2}$ & $0.72_{-0.03}^{+0.02}$ 
& $3.86_{-0.71}^{+0.73}$ & & $70.7_{-1.9}^{+1.9}$ & 
$0.73_{-0.02}^{+0.02}$ & $\leq 0.393$ \\ 
\vspace{0.1cm} TDSL+Planck (1) & $71.0_{-2.0}^{+2.0}$ & 
$0.71_{-0.01}^{+0.01}$ & $3.45_{-0.24}^{+0.23}$ & & $68.1_{-1.2}^{+1.1}$ 
& $0.70_{-0.02}^{+0.01}$ & $\leq 0.199$ \\ 
\vspace{0.1cm} (1)+BAO & $69.6_{-1.3}^{+1.4}$ & $0.70_{-0.01}^{+0.01}$ & 
$3.34_{-0.21}^{+0.21}$ & & $67.9_{-0.6}^{+0.6}$ & $0.69_{-0.01}^{+0.01}$ 
& $\leq 0.182$ \\ 
 (1)+BAO+CMBL &   &   &   & & $67.9_{-0.7}^{+0.6}$ & 
$0.69_{-0.01}^{+0.01}$ & $\leq 0.216$  \\ 
\hline
 \end{tabular}

 \vspace{0.5cm}

 \begin{tabular}{l c c c c c c c c c}
  \hline
  & \multicolumn{4}{c}{\nnumnuLCDM} & & \multicolumn{4}{c}{\owCDM}  \\
  & \hc & $\Omega_{\Lambda}$ & \nnu & \mnu (eV) & & \hc & $\Ode$ & $\Ok$
& $w$ \\
  \hline
\vspace{0.1cm} TDSL+Planck (1) & $70.8_{-2.1}^{+2.0}$ & 
$0.71_{-0.02}^{+0.02}$ & $3.44_{-0.24}^{+0.24}$ & $\leq0.274$ & & 
$88.4_{-7.2}^{+5.9}$ & $0.83_{-0.03}^{+0.02}$ & 
$-0.010_{-0.003}^{+0.003}$ & $-2.10_{-0.41}^{+0.34}$ \\ 
\vspace{0.1cm} (1)+CMBL & $70.8_{-2.1}^{+2.1}$ & $0.71_{-0.02}^{+0.02}$ 
& $3.44_{-0.24}^{+0.25}$ & $\leq0.347$ & & $77.9_{-4.2}^{+5.0}$ & 
$0.77_{-0.03}^{+0.03}$ & $-0.003_{-0.004}^{+0.004}$ & 
$-1.37_{-0.23}^{+0.18}$ \\ 
\vspace{0.1cm} (1)+BAO+CMBL &   &   &   &   & & $70.0_{-1.7}^{+2.1}$ & 
$0.71_{-0.02}^{+0.02}$ & $-0.000_{-0.003}^{+0.004}$ & 
$-1.07_{-0.10}^{+0.09}$ \\ 
\hline
\end{tabular}
 \label{tab:cosmoparams}
\end{table*}

\subsubsection{One-parameter extensions}\label{sec:cosmo_oneparamext}

We first consider one-parameter extensions to the standard model, where
we relax the constraints on one additional cosmological parameter from
flat \LCDM. We present in Figure~\ref{fig:cosmo_oneparamext} the
two-dimensional marginalized parameter space for a selection of cosmological models
for which the impact of TDSL is the most meaningful.

In the \oLCDM model, we consider a non-flat Universe with $\Ok \neq 0$.
In the \nnuLCDM model, we consider a variable effective number
of relativistic neutrino species \nnu\ with a fixed total mass of
neutrinos \mnu\ $=0.06$ eV. In the \mnuLCDM model we consider a
variable \mnu\ with a fixed \nnu\ $=3.046$. Finally, in the \wCDM model
we consider a time-invariant dark energy equation of state $w$. A
detailed description of these models is given in
Table~\ref{tab:cosmologies}.

For each probe, or combination of probes, we draw
the 95\% credible region contours as colored lines. When combined with TDSL,
the updated
credible
region is displayed as a filled area. When
importance sampling using priors based on other probes, it is important to
verify that the respective constraints in the parameter space overlap.
If they do not, the probes considered may not be efficiently combined.
With this in mind, we plot in
each cosmology the 95\% credible region for TDSL only (and uniform priors) as
thin solid black lines.
We note that in all one-parameter extensions presented here, the 2D
marginalized TDSL and Planck 95\% credible regions at least partially
overlap, although in the \oLCDM\ and \mnuLCDM\ cosmologies, the 1D
marginalized posterior mean value for \hc\ from TDSL lies outside the
95\% credible region of Planck.
We consider the
overlaps to be sufficient to justify our importance sampling TDSL
with Planck, but emphasize that the joint constraints must
be interpreted cautiously. WMAP and Planck constraints are in agreement
with each other, this being at least partly due to the large
parameter space covered in the credible region of WMAP. This also
results in a much wider overlap with the TDSL 95\% credible regions.

We summarize our inferred values for \hc and other cosmological
parameters of each cosmology in Table~\ref{tab:cosmoparams}. In the
\oLCDM cosmology, both WMAP+TDSL and Planck+TDSL are consistent with a
flat
universe. The constraints on $\Ok$ from Planck+TDSL are approximately
twice as large as those from Planck+BAO+JLA. In the \mnuLCDM cosmology, the upper
bound of the sum of the neutrino masses \mnu from WMAP+TDSL is
approximately twice as large as the prediction from Planck+TDSL. The
addition of TDSL lowers the upper bound from Planck+BAO by about 5\%. The joint
constraint from Planck+BAO+TDSL yields \mnu$\leq0.182$~eV with 95\%
probability. In the \nnuLCDM cosmology, both WMAP+TDSL and
Planck+TDSL suggest an effective number of relativistic neutrino
species
higher than the standard cosmological prediction of \nnu$=3.046$. The
Planck+TDSL value is similar in precision to Planck+BAO, yet the two
values are in
mild tension, the former being 1.3$\sigma$ higher. The combination of
Planck+BAO+TDSL yields \nnu$=3.34\pm0.21$, also in mild tension with
the
standard cosmological prediction. In the \wCDM cosmology, Planck+TDSL
points towards $w=-1.38_{-0.16}^{+0.14}$, a result in tension with a
cosmological constant ($w=-1$) at a 2.3$\sigma$ level. This value is
lower than other values for phantom dark energy reported
in the literature \citep[see e.g.][]{Freedman2012, Collett2014}.
With WMAP+TDSL we find  $w=-1.24_{-0.20}^{+0.16}$, consistent with the
previous measurement from our group using just \blens\ and \rxjlens\
combined with WMAP, of \citep[$w=-1.14_{-0.20}^{+0.17}$;][]{Suyu2013}.

\subsubsection{Two-parameter extensions}

We now consider cosmological models where we relax the
priors on two cosmological parameters from flat \LCDM. Following
the discussion of Section~\ref{sec:cosmo_oneparamext} where we noted that
the individual TDSL and
Planck 95\% credible regions only partially overlap, we consider
here two cosmological models that reduce the tension between these two
probes. First, we consider the
\nnumnuLCDM model, where both the effective number of relativistic
neutrino species \nnu as well as their total mass \mnu are allowed to
vary. Second, we consider the \owCDM model where we relax the
constraints on both the curvature, $\Ok$, and the dark energy equation
of state parameter $w$ simultaneously. For the \owCDM
model the Planck team does not publicly provide MCMC chains. We
therefore generate additional chains using the publicly available
Planck cosmological likelihood function, {\it plik}
\citep{PlanckTopology2015}. Temperature power spectra were computed
using the Cosmic Linear Anisotropy Solving System Boltzman code
\citep[CLASS;][]{Blas2011, Lesgourgues2014} and MCMC chains were
generated with
the MontePython sampler \citep{Audren2013}.

Figure~\ref{fig:cosmo_twoparamext} presents the two-dimensional marginalized
95\% credible regions for these two models, and the
bottom of
Table~\ref{tab:cosmoparams} reports the 1D marginalized posterior median values and 1$\sigma$
uncertainties of the corresponding model parameters.
We note that this time, the TDSL and Planck 95\% credible regions are in
much better agreement than in the one-parameter extension models. In
the \nnumnuLCDM cosmology, Planck alone and Planck+CMBL are in
agreement with the standard cosmological prediction of \nnu, yet the
constraints are rather large. When adding TDSL, the constraints are
strongly tightened and we obtain \nnu$=3.44\pm0.24$, in mild tension
with the standard cosmological prediction of $\Neff=3.046$. Similarly, the constraints on
the maximum neutrino mass are tightened by a factor $\simeq$3 when adding
TDSL, yielding \mnu$\leq0.274$~eV with 95\% probability.
In the \owCDM cosmology,
Planck+CMBL+TDSL
yields $\Ok=-0.003^{+0.004}_{-0.004}$, in good agreement with
Planck+CMBL+BAO and
in favor of a flat universe. However, a tension in the dark
energy
equation of state $w$ still remains, as Planck+CMBL+TDSL yields
$w=-1.37^{-0.23}_{+0.18}$, $2\sigma$
lower than the cosmological constant prediction.

%=========================================================
%			CONCLUSIONS
%=========================================================

\section{Conclusions}
\label{sec:conclusion}

Using multiple telescopes in the Southern and Northern hemispheres, we
have monitored the quadruple imaged strong gravitational lens \hequad\ for 13
years with an average cadence of one observing epoch every 3.6 days. We
analyse the imaging data using the MCS deconvolution algorithm
\citep{Magain1998} on a total of 876 observing epochs to produce the
light curves of the four lensed images, with an rms photometric
precision of 10 mmag on the brightest quasar image.

We measured the time delays between each pair of lensed images using the
free-knot spline technique and the regression difference
technique from the {\tt PyCS} package \citep{Tewes2013a}. Our
uncertainty estimation involves the generation of synthetic light curves
that closely mimic the intrinsic and extrinsic features of the real
data. To test the robustness of our measurements, we vary parameters 
such as the number of knots in the splines, the initial parameters
used for the deconvolution photometry, and the length of the considered light curves.
The two curve shifting techniques agree well with each other
both on the point estimation of the delays and on the estimated
uncertainty. The smallest relative uncertainty, of 6.5\%, is obtained for the
A-D pair of images. For this pair involving image A, our present measurement 
is twice as precise as the earlier result by \citet{Courbin2011}.

In \PaperIV, we use our new COSMOGRAIL time delays for \hequad\
to compute its time-delay distance. Very importantly, this is done in a
blind way with respect to the inference of cosmological parameters. In
this paper, we combine the time-delay distance likelihoods from
\hequad\ with the published
ones from \blens\ and \rxjlens\ to create a Time Delays Strong Lensing
(TDSL) probe.
We also combine the latter with other cosmological probes such as WMAP,
Planck, BAO and JLA to constrain cosmological parameters for a large
range of extended cosmological models. Our main conclusions are as
follows:

\begin{itemize}

 \item TDSL alone is weakly sensitive to the matter density, $\Om$,
 curvature, $\Ok$ and dark energy density $\Ode$ and equation of state 
$w$. Its
 primary sensitivity to \hc allows us to break degeneracies of CMB
 probes in extended cosmological models.

 \item In a flat \LCDM cosmology with uniform priors on \hc and
$\Om$, TDSL
alone yields \hc$=71.9_{-3.0}^{+2.4}\ {\rm km\, s^{-1}\,Mpc^{-1}}$.
When enforcing $\Om$=0.32 from the most recent Planck
 results, we find \hc=$72.8\pm2.4\
 {\rm km\, s^{-1}\,Mpc^{-1}}$. These results are in excellent agreement
with the most recent measurements using the distance
 ladder, but are in tension with the CMB measurements from Planck.

 \item In a non-flat \LCDM cosmology, we find, using TDSL and Planck,
 \hc=$69.2_{-2.2}^{+1.4}\ {\rm km\, s^{-1}\,Mpc^{-1}}$ and
$\Ok=0.003_{-0.006}^{+0.004}$, in
 agreement with a flat universe.

 \item In a flat \wCDM cosmology in combination with Planck, we find a
 2.3$\sigma$ tension from a cosmological constant in favor of a phantom
 form of dark energy. Our joint constraints in this cosmology are
 \hc$=79.0_{-4.2}^{+4.4}\ {\rm km\, s^{-1}\,Mpc^{-1}}$,
$\Ode=0.77_{-0.03}^{+0.02}$ and
 $w=-1.38_{-0.16}^{+0.14}$.

 \item In a flat \mnuLCDM cosmology, in combination with Planck and BAO 
we
 tighten the constraints on the maximum mass of neutrinos to
 \mnu$\leq0.182$~eV, while removing the tension in \hc.

 \item In a flat \nnuLCDM cosmology, in combination with Planck and BAO
 we find \nnu$=3.34\pm0.21$, i.e. 1.3$\sigma$
 higher than the standard cosmological value.
 This mild tension remains
 when the constraints on both \mnu and \nnu are relaxed.
 
 \item In a \owCDM cosmology, in combination with Planck and CMBL, we
find
 \hc$=77.9_{-4.2}^{+5.0}\ {\rm km\, s^{-1}\,Mpc^{-1}}$,
$\Ode=0.77_{-0.03}^{+0.03}$,
 $\Ok=-0.003_{-0.004}^{+0.004}$ and
 $w=-1.37_{-0.23}^{+0.18}$. Similarly to the \oLCDM and \wCDM
 cosmologies, we are in good agreement with a flat universe and 
in tension with a cosmological constant, respectively.

\end{itemize}

We emphasize that despite reporting parameter 
constraints for a large variety of cosmological models beyond \LCDM,
we choose not to comment on whether a particular model is favored over 
the others. Such 
an exercise would require a well motivated choice of priors for these 
models, which is not within the scope of this work.\\

The combined strengths of our H0LiCOW lens modeling and COSMOGRAIL
monitoring indicate that quasar time-delay cosmography is now a mature
field, producing precise and accurate inferences of cosmological 
parameters, that are independent of any other cosmological
probe. Still, our results can be improved in at least 
four ways:

\begin{enumerate}

\item Continuing to enlarge the sample. Two more objects with excellent 
time-delay measurements as well as high-resolution
imaging and spectroscopic data remain to be analysed in the H0LiCOW project
(see \PaperI). When completed, H0LiCOW is expected to
provide a measurement of \hc to better than 3.5\% in a non-flat \LCDM universe with
flat priors on $\Om$ and $\OL$. Data of quality comparable
to
those obtained for H0LiCOW are in the process of being obtained for
another four systems with measured time delays from COSMOGRAIL
(HST-GO-14254; PI: Treu).
Meanwhile, current and planned wide field imaging surveys such
as DES, KiDS, HSC, LSST, Euclid and WFIRST, should discover hundreds of
new gravitational lens systems suitable for time-delay cosmography
\citep{O+M10}. For example, the dedicated search in the Dark Energy
Survey
STRIDES\footnote{strides.astro.ucla.edu} has already confirmed two new
lenses from the Year1 data
\citep{Agnello2016}.

\item Improve the lens modeling accuracy. The tests carried out in our current (\PaperIV)
and past work \citep{Suyu2014}, the good internal agreement between
the three measured systems (Section \ref{sec:ubgcosmo}), and 
independent analysis
based on completely independent codes \citep{Birrer2016}, show that
our lens models are sufficiently complex given the currently available
data. However, as the number of systems being analysed  grows,
random uncertainties in the cosmological parameters will fall, and residual systematic uncertainties
related to degeneracies inherent to gravitational lensing will need to
be investigated in more detail. Following the work of
\citet{Xu2016}, detailed hydro N-body simulations of lensing galaxies
in combination with ray-shooting can be used to evaluate the impact of
the lensing degeneracies on cosmological results in view of future
observations with the JWST or 30-m class ground-based telescopes with
adaptive optics, and to drive development of improved lens modeling
techniques and assumptions appropriate to the density structures we expect.

\item Improve the absolute mass calibration. Spatially resolved 2D kinematics of the lens
galaxies, to be obtained either with JWST and with integral field
spectrographs mounted on large ground-based telescopes with adaptive
optics, should further improve both the precision for each system and our
ability to test for residual systematics, including those arising from
the mass sheet and source position transformations 
\citep{Schneider2013, Schneider2014, Unruh2016}. The same data should 
also allow
us to use constraints from the stellar mass or mass profile of lens 
galaxies as attempted in \citet{Courbin2011} with slit spectroscopy. 
Alternatively, the 
mass-sheet degeneracy can be lifted if the absolute 
luminosity of the source is known \citep{Falco1985}, which is the case 
for lensed standard candles \citep[see e.g.][that report the first 
discovery of a lensed type-Ia Supernovae]{Goobar2016}. However, such 
configurations happens far less often than lensed quasars.

\item  Measuring time delays with the current photometric precision and
time sampling of monitoring data requires long and time-consuming
campaigns, and is currently not possible for hundreds of objects. Increasing the
monitoring efficiency is possible, by catching extremely small (mmag)
and fast (days) variations in the quasar light curves. Such data can
be obtained with daily observations with 2-m class telescopes in good
seeing conditions, a project that will be implemented in the context
of the extended COSMOGRAIL program (eCOSMOGRAIL; Courbin et al. 2016,
in prep) to measure quasar time delays in only 1 or 2 observing
seasons. Furthermore, in the long run, LSST should be able to provide
sufficiently accurate time delays for hundreds of systems from the
survey data itself \citep{Liao2015}, and enable sub-percent precision
on \hc in the next decade \citep{Treu2016}.

\end{enumerate}

%-------------------------------------------------------------------------------

\section*{Acknowledgments}
We thank Adriano Agnello, Roger Blandford, Geoff ChihFan Chen, Xuheng 
Ding, Yashar Hezaveh, Kai Liao, John McKean, Danka 
Paraficz, Olga Tihhonova, and Simona Vegetti for their contributions to 
the H0LiCOW project.
We thank the anonymous referee for his or her comments.
We also thank all the observers at the Euler, SMARTS, Mercator
and Maidanak telescopes who participated in the queue-mode
observations. H0LiCOW and COSMOGRAIL are made possible thanks to the continuous work of all observers and technical staff obtaining the monitoring observations, in particular at the Swiss Euler telescope at La Silla Observatory. The Euler telescope is supported by the Swiss National Science Foundation.
We are grateful to Thomas Tram for expert help in
implementing CLASS and MontePython to create MCMC chains for
non-standard cosmologies. Numerical computations for the CMB likelihoods
were done on the Sciama High Performance Compute (HPC) cluster which is
supported by the ICG, SEPNet and the University of Portsmouth
V.B., F.C. and G.M. acknowledge the support of the Swiss National
Science Foundation (SNSF).
S.H.S. acknowledges support from the Max Planck Society through the Max
Planck Research Group.  This work is supported in part by the Ministry
of Science and Technology in Taiwan via grant
MOST-103-2112-M-001-003-MY3.
K.C.W. is supported by an EACOA Fellowship awarded by the East Asia Core
Observatories Association, which consists of the Academia Sinica
Institute of Astronomy and Astrophysics, the National Astronomical
Observatory of Japan, the National Astronomical Observatories of the
Chinese Academy of Sciences, and the Korea Astronomy and Space Science
Institute.
P.J.M. acknowledges support from the U.S.\ Department of Energy under
contract number DE-AC02-76SF00515.
C.E.R and C.D.F. were funded through the NSF grant AST-1312329,
``Collaborative Research: Accurate cosmology with strong gravitational
lens time delays,'' and the {\it HST} grant GO-12889.
D.S. acknowledges funding support from a {\it {Back to Belgium}} grant
from the Belgian Federal Science Policy (BELSPO).
M.T. acknowledges support by a fellowship of the Alexander von Humboldt
Foundation and the DFG grant Hi 1495/2-1.
S.H. acknowledges support by the DFG cluster of excellence \lq{}Origin
and Structure of the Universe\rq{}
(\href{http://www.universe-cluster.de}{\texttt{www.universe-cluster.de}}).
T.T. thanks the Packard Foundation for generous support through a
Packard Research Fellowship, the
NSF for funding through NSF grant AST-1450141, ``Collaborative Research:
Accurate cosmology with strong gravitational lens time delays''.
L.V.E.K is supported in part through an NWO-VICI career grant (project
number 639.043.308).

Based on observations made with the NASA/ESA Hubble Space Telescope,
obtained at the Space Telescope Science Institute, which is operated by
the Association of Universities for Research in Astronomy, Inc., under
NASA contract NAS 5-26555. These observations are associated with
program \#12889, \#10158, \#9744 and \#7422. Support for program
\#12889 was provided by NASA
through a grant from the Space Telescope Science Institute, which is
operated by the Association of Universities for Research in Astronomy,
Inc., under NASA contract NAS 5-26555.

%-------------------------------------------------------------------------------
% bibliography:

\bibliography{he0435delays_cosmo_revisionv1}

\begin{thebibliography}{}
\makeatletter
\relax
\def\mn@urlcharsother{\let\do\@makeother \do\$\do\&\do\#\do\^\do\_\do\%\do\~}
\def\mn@doi{\begingroup\mn@urlcharsother \@ifnextchar [ {\mn@doi@}
  {\mn@doi@[]}}
\def\mn@doi@[#1]#2{\def\@tempa{#1}\ifx\@tempa\@empty \href
  {http://dx.doi.org/#2} {doi:#2}\else \href {http://dx.doi.org/#2} {#1}\fi
  \endgroup}
\def\mn@eprint#1#2{\mn@eprint@#1:#2::\@nil}
\def\mn@eprint@arXiv#1{\href {http://arxiv.org/abs/#1} {{\tt arXiv:#1}}}
\def\mn@eprint@dblp#1{\href {http://dblp.uni-trier.de/rec/bibtex/#1.xml}
  {dblp:#1}}
\def\mn@eprint@#1:#2:#3:#4\@nil{\def\@tempa {#1}\def\@tempb {#2}\def\@tempc
  {#3}\ifx \@tempc \@empty \let \@tempc \@tempb \let \@tempb \@tempa \fi \ifx
  \@tempb \@empty \def\@tempb {arXiv}\fi \@ifundefined
  {mn@eprint@\@tempb}{\@tempb:\@tempc}{\expandafter \expandafter \csname
  mn@eprint@\@tempb\endcsname \expandafter{\@tempc}}}

\bibitem[\protect\citeauthoryear{{Addison}, {Huang}, {Watts}, {Bennett},
  {Halpern}, {Hinshaw}  \& {Weiland}}{{Addison} et~al.}{2016}]{Addison2016}
{Addison} G.~E.,  {Huang} Y.,  {Watts} D.~J.,  {Bennett} C.~L.,  {Halpern} M.,
  {Hinshaw} G.,   {Weiland} J.~L.,  2016, \mn@doi [\apj]
  {10.3847/0004-637X/818/2/132}, \href
  {http://adsabs.harvard.edu/abs/2016ApJ...818..132A} {818, 132}

\bibitem[\protect\citeauthoryear{{Agnello}, {Sonnenfeld}, {Suyu}, {Treu},
  {Fassnacht}, {Mason}, {Brada{\v c}}  \& {Auger}}{{Agnello}
  et~al.}{2016}]{Agnello2016}
{Agnello} A.,  {Sonnenfeld} A.,  {Suyu} S.~H.,  {Treu} T.,  {Fassnacht} C.~D.,
  {Mason} C.,  {Brada{\v c}} M.,   {Auger} M.~W.,  2016, \mn@doi [\mnras]
  {10.1093/mnras/stw529}, \href
  {http://adsabs.harvard.edu/abs/2016MNRAS.458.3830A} {458, 3830}

\bibitem[\protect\citeauthoryear{{Anderson} et~al.,}{{Anderson}
  et~al.}{2012}]{Anderson2012}
{Anderson} L.,  et~al., 2012, \mn@doi [\mnras]
  {10.1111/j.1365-2966.2012.22066.x}, \href
  {http://adsabs.harvard.edu/abs/2012MNRAS.427.3435A} {427, 3435}

\bibitem[\protect\citeauthoryear{{Audren}, {Lesgourgues}, {Benabed}  \&
  {Prunet}}{{Audren} et~al.}{2013}]{Audren2013}
{Audren} B.,  {Lesgourgues} J.,  {Benabed} K.,   {Prunet} S.,  2013, \mn@doi
  [\jcap] {10.1088/1475-7516/2013/02/001}, \href
  {http://adsabs.harvard.edu/abs/2013JCAP...02..001A} {2, 001}

\bibitem[\protect\citeauthoryear{{Bennett} et~al.,}{{Bennett}
  et~al.}{2013}]{Bennett2013}
{Bennett} C.~L.,  et~al., 2013, \mn@doi [\apjs] {10.1088/0067-0049/208/2/20},
  \href {http://adsabs.harvard.edu/abs/2013ApJS..208...20B} {208, 20}

\bibitem[\protect\citeauthoryear{{Bertin} \& {Arnouts}}{{Bertin} \&
  {Arnouts}}{1996}]{Bertin1996}
{Bertin} E.,  {Arnouts} S.,  1996, \mn@doi [\aaps] {10.1051/aas:1996164}, \href
  {http://adsabs.harvard.edu/abs/1996A%26AS..117..393B} {117, 393}

\bibitem[\protect\citeauthoryear{{Betoule} et~al.,}{{Betoule}
  et~al.}{2013}]{Betoule2013}
{Betoule} M.,  et~al., 2013, \mn@doi [\aap] {10.1051/0004-6361/201220610},
  \href {http://adsabs.harvard.edu/abs/2013A%26A...552A.124B} {552, A124}

\bibitem[\protect\citeauthoryear{{Beutler} et~al.,}{{Beutler}
  et~al.}{2011}]{Beutler2011}
{Beutler} F.,  et~al., 2011, \mn@doi [\mnras]
  {10.1111/j.1365-2966.2011.19250.x}, \href
  {http://adsabs.harvard.edu/abs/2011MNRAS.416.3017B} {416, 3017}

\bibitem[\protect\citeauthoryear{{Birrer}, {Amara}  \& {Refregier}}{{Birrer}
  et~al.}{2015}]{Birrer2016}
{Birrer} S.,  {Amara} A.,   {Refregier} A.,  2015, preprint, \href
  {http://adsabs.harvard.edu/abs/2015arXiv151103662B} {} (\mn@eprint {arXiv}
  {1511.03662})

\bibitem[\protect\citeauthoryear{{Blackburne}, {Kochanek}, {Chen}, {Dai}  \&
  {Chartas}}{{Blackburne} et~al.}{2014}]{Blackburne2014}
{Blackburne} J.~A.,  {Kochanek} C.~S.,  {Chen} B.,  {Dai} X.,   {Chartas} G.,
  2014, \mn@doi [\apj] {10.1088/0004-637X/789/2/125}, \href
  {http://adsabs.harvard.edu/abs/2014ApJ...789..125B} {789, 125}

\bibitem[\protect\citeauthoryear{{Blake} et~al.,}{{Blake}
  et~al.}{2011}]{Blake2011}
{Blake} C.,  et~al., 2011, \mn@doi [\mnras] {10.1111/j.1365-2966.2011.19592.x},
  \href {http://adsabs.harvard.edu/abs/2011MNRAS.418.1707B} {418, 1707}

\bibitem[\protect\citeauthoryear{{Blas}, {Lesgourgues}  \& {Tram}}{{Blas}
  et~al.}{2011}]{Blas2011}
{Blas} D.,  {Lesgourgues} J.,   {Tram} T.,  2011, {CLASS: Cosmic Linear
  Anisotropy Solving System}, Astrophysics Source Code Library (\mn@eprint
  {ascl} {1106.020})

\bibitem[\protect\citeauthoryear{{Bonamente}, {Joy}, {LaRoque}, {Carlstrom},
  {Reese}  \& {Dawson}}{{Bonamente} et~al.}{2006}]{Bonamente2006}
{Bonamente} M.,  {Joy} M.~K.,  {LaRoque} S.~J.,  {Carlstrom} J.~E.,  {Reese}
  E.~D.,   {Dawson} K.~S.,  2006, \mn@doi [\apj] {10.1086/505291}, \href
  {http://adsabs.harvard.edu/abs/2006ApJ...647...25B} {647, 25}

\bibitem[\protect\citeauthoryear{{Bonvin}, {Tewes}, {Courbin}, {Kuntzer},
  {Sluse}  \& {Meylan}}{{Bonvin} et~al.}{2016}]{Bonvin2016}
{Bonvin} V.,  {Tewes} M.,  {Courbin} F.,  {Kuntzer} T.,  {Sluse} D.,   {Meylan}
  G.,  2016, \mn@doi [Astronomy and Astrophysics]
  {10.1051/0004-6361/201526704}, \href
  {http://adsabs.harvard.edu/abs/2016A%26A...585A..88B} {585, A88}

\bibitem[\protect\citeauthoryear{{Braibant}, {Hutsem{\'e}kers}, {Sluse},
  {Anguita}  \& {Garc{\'{\i}}a-Vergara}}{{Braibant}
  et~al.}{2014}]{Braibant2014}
{Braibant} L.,  {Hutsem{\'e}kers} D.,  {Sluse} D.,  {Anguita} T.,
  {Garc{\'{\i}}a-Vergara} C.~J.,  2014, \mn@doi [\aap]
  {10.1051/0004-6361/201423633}, \href
  {http://adsabs.harvard.edu/abs/2014A%26A...565L..11B} {565, L11}

\bibitem[\protect\citeauthoryear{{Cantale}, {Courbin}, {Tewes}, {Jablonka.}  \&
  {Meylan}}{{Cantale} et~al.}{2016}]{Cantale2016a}
{Cantale} N.,  {Courbin} F.,  {Tewes} M.,  {Jablonka.} P.,   {Meylan} G.,
  2016, ArXiv1602.02167, \href
  {http://adsabs.harvard.edu/abs/2016arXiv160202167C} {}

\bibitem[\protect\citeauthoryear{{Chen} et~al.,}{{Chen}
  et~al.}{2016}]{Chen2016}
{Chen} G.~C.~F.,  et~al., 2016, ArXiv1601.01321, \href
  {http://adsabs.harvard.edu/abs/2016arXiv160101321C} {}

\bibitem[\protect\citeauthoryear{{Collett} \& {Auger}}{{Collett} \&
  {Auger}}{2014}]{Collett2014}
{Collett} T.~E.,  {Auger} M.~W.,  2014, \mn@doi [\mnras]
  {10.1093/mnras/stu1190}, \href
  {http://adsabs.harvard.edu/abs/2014MNRAS.443..969C} {443, 969}

\bibitem[\protect\citeauthoryear{{Collett} et~al.,}{{Collett}
  et~al.}{2013}]{Collett2013}
{Collett} T.~E.,  et~al., 2013, \mn@doi [\mnras] {10.1093/mnras/stt504}, \href
  {http://adsabs.harvard.edu/abs/2013MNRAS.432..679C} {432, 679}

\bibitem[\protect\citeauthoryear{{Courbin}, {Eigenbrod}, {Vuissoz}, {Meylan}
  \& {Magain}}{{Courbin} et~al.}{2005}]{Courbin2005}
{Courbin} F.,  {Eigenbrod} A.,  {Vuissoz} C.,  {Meylan} G.,   {Magain} P.,
  2005, in {Mellier} Y.,  {Meylan} G.,  eds,  IAU Symposium Vol. 225,
  Gravitational Lensing Impact on Cosmology. pp 297--303,
  \mn@doi{10.1017/S1743921305002097}

\bibitem[\protect\citeauthoryear{{Courbin} et~al.,}{{Courbin}
  et~al.}{2011}]{Courbin2011}
{Courbin} F.,  et~al., 2011, \mn@doi [Astronomy and Astrophysics]
  {10.1051/0004-6361/201015709}, \href
  {http://adsabs.harvard.edu/abs/2011A%26A...536A..53C} {536, A53}

\bibitem[\protect\citeauthoryear{{Di Valentino}, {Melchiorri}  \& {Silk}}{{Di
  Valentino} et~al.}{2016}]{DiValentino2016}
{Di Valentino} E.,  {Melchiorri} A.,   {Silk} J.,  2016, ArXiv1606.00634, \href
  {http://adsabs.harvard.edu/abs/2016arXiv160600634D} {}

\bibitem[\protect\citeauthoryear{{Dobler}, {Fassnacht}, {Treu}, {Marshall},
  {Liao}, {Hojjati}, {Linder}  \& {Rumbaugh}}{{Dobler}
  et~al.}{2015}]{Dobler2015}
{Dobler} G.,  {Fassnacht} C.~D.,  {Treu} T.,  {Marshall} P.,  {Liao} K.,
  {Hojjati} A.,  {Linder} E.,   {Rumbaugh} N.,  2015, \mn@doi [\apj]
  {10.1088/0004-637X/799/2/168}, \href
  {http://adsabs.harvard.edu/abs/2015ApJ...799..168D} {799, 168}

\bibitem[\protect\citeauthoryear{{Efstathiou}}{{Efstathiou}}{2014}]{Efstathiou2014}
{Efstathiou} G.,  2014, \mn@doi [\mnras] {10.1093/mnras/stu278}, \href
  {http://adsabs.harvard.edu/abs/2014MNRAS.440.1138E} {440, 1138}

\bibitem[\protect\citeauthoryear{{Eigenbrod}, {Courbin}, {Dye}, {Meylan},
  {Sluse}, {Vuissoz}  \& {Magain}}{{Eigenbrod} et~al.}{2006a}]{Eigenbrod2006a}
{Eigenbrod} A.,  {Courbin} F.,  {Dye} S.,  {Meylan} G.,  {Sluse} D.,  {Vuissoz}
  C.,   {Magain} P.,  2006a, \mn@doi [\aap] {10.1051/0004-6361:20054423}, \href
  {http://adsabs.harvard.edu/abs/2006A%26A...451..747E} {451, 747}

\bibitem[\protect\citeauthoryear{{Eigenbrod}, {Courbin}, {Meylan}, {Vuissoz}
  \& {Magain}}{{Eigenbrod} et~al.}{2006b}]{Eigenbrod2006b}
{Eigenbrod} A.,  {Courbin} F.,  {Meylan} G.,  {Vuissoz} C.,   {Magain} P.,
  2006b, \mn@doi [\aap] {10.1051/0004-6361:20054454}, \href
  {http://adsabs.harvard.edu/abs/2006A%26A...451..759E} {451, 759}

\bibitem[\protect\citeauthoryear{{Eulaers} et~al.,}{{Eulaers}
  et~al.}{2013}]{Eulaers2013}
{Eulaers} E.,  et~al., 2013, \mn@doi [\aap] {10.1051/0004-6361/201321140},
  \href {http://adsabs.harvard.edu/abs/2013A%26A...553A.121E} {553, A121}

\bibitem[\protect\citeauthoryear{{Falco}, {Gorenstein}  \& {Shapiro}}{{Falco}
  et~al.}{1985}]{Falco1985}
{Falco} E.~E.,  {Gorenstein} M.~V.,   {Shapiro} I.~I.,  1985, \mn@doi [\apjl]
  {10.1086/184422}, \href {http://adsabs.harvard.edu/abs/1985ApJ...289L...1F}
  {289, L1}

\bibitem[\protect\citeauthoryear{{Fassnacht}, {Xanthopoulos}, {Koopmans}  \&
  {Rusin}}{{Fassnacht} et~al.}{2002}]{Fassnacht2002}
{Fassnacht} C.~D.,  {Xanthopoulos} E.,  {Koopmans} L.~V.~E.,   {Rusin} D.,
  2002, \mn@doi [\apj] {10.1086/344368}, \href
  {http://adsabs.harvard.edu/abs/2002ApJ...581..823F} {581, 823}

\bibitem[\protect\citeauthoryear{{Fassnacht}, {Gal}, {Lubin}, {McKean},
  {Squires}  \& {Readhead}}{{Fassnacht} et~al.}{2006}]{Fassnacht2006}
{Fassnacht} C.~D.,  {Gal} R.~R.,  {Lubin} L.~M.,  {McKean} J.~P.,  {Squires}
  G.~K.,   {Readhead} A.~C.~S.,  2006, \mn@doi [\apj] {10.1086/500927}, \href
  {http://adsabs.harvard.edu/abs/2006ApJ...642...30F} {642, 30}

\bibitem[\protect\citeauthoryear{{Fassnacht}, {Koopmans}  \&
  {Wong}}{{Fassnacht} et~al.}{2011}]{Fassnacht2011}
{Fassnacht} C.~D.,  {Koopmans} L.~V.~E.,   {Wong} K.~C.,  2011, \mn@doi
  [\mnras] {10.1111/j.1365-2966.2010.17591.x}, \href
  {http://adsabs.harvard.edu/abs/2011MNRAS.410.2167F} {410, 2167}

\bibitem[\protect\citeauthoryear{{Freedman}, {Madore}, {Scowcroft}, {Burns},
  {Monson}, {Persson}, {Seibert}  \& {Rigby}}{{Freedman}
  et~al.}{2012}]{Freedman2012}
{Freedman} W.~L.,  {Madore} B.~F.,  {Scowcroft} V.,  {Burns} C.,  {Monson} A.,
  {Persson} S.~E.,  {Seibert} M.,   {Rigby} J.,  2012, \mn@doi [\apj]
  {10.1088/0004-637X/758/1/24}, \href
  {http://adsabs.harvard.edu/abs/2012ApJ...758...24F} {758, 24}

\bibitem[\protect\citeauthoryear{{Gao} et~al.,}{{Gao} et~al.}{2016}]{Gao2016}
{Gao} F.,  et~al., 2016, \mn@doi [\apj] {10.3847/0004-637X/817/2/128}, \href
  {http://adsabs.harvard.edu/abs/2016ApJ...817..128G} {817, 128}

\bibitem[\protect\citeauthoryear{{Giusarma}, {Gerbino}, {Mena}, {Vagnozzi},
  {Ho}  \& {Freese}}{{Giusarma} et~al.}{2016}]{Giusarma2016}
{Giusarma} E.,  {Gerbino} M.,  {Mena} O.,  {Vagnozzi} S.,  {Ho} S.,   {Freese}
  K.,  2016, ArXiv1605.04320, \href
  {http://adsabs.harvard.edu/abs/2016arXiv160504320G} {}

\bibitem[\protect\citeauthoryear{{Goobar} et~al.,}{{Goobar}
  et~al.}{2016}]{Goobar2016}
{Goobar} A.,  et~al., 2016, ArXiv1611.00014, \href
  {http://adsabs.harvard.edu/abs/2016arXiv161100014G} {}

\bibitem[\protect\citeauthoryear{{Greene} et~al.,}{{Greene}
  et~al.}{2013}]{Greene2013}
{Greene} Z.~S.,  et~al., 2013, \mn@doi [\apj] {10.1088/0004-637X/768/1/39},
  \href {http://adsabs.harvard.edu/abs/2013ApJ...768...39G} {768, 39}

\bibitem[\protect\citeauthoryear{{Heavens}, {Jimenez}  \& {Verde}}{{Heavens}
  et~al.}{2014}]{Heavens2014}
{Heavens} A.,  {Jimenez} R.,   {Verde} L.,  2014, \mn@doi [Physical Review
  Letters] {10.1103/PhysRevLett.113.241302}, \href
  {http://adsabs.harvard.edu/abs/2014PhRvL.113x1302H} {113, 241302}

\bibitem[\protect\citeauthoryear{{Heymans} et~al.,}{{Heymans}
  et~al.}{2012}]{Heymans2012}
{Heymans} C.,  et~al., 2012, \mn@doi [\mnras]
  {10.1111/j.1365-2966.2012.21952.x}, \href
  {http://adsabs.harvard.edu/abs/2012MNRAS.427..146H} {427, 146}

\bibitem[\protect\citeauthoryear{{Hilbert}, {White}, {Hartlap}  \&
  {Schneider}}{{Hilbert} et~al.}{2007}]{Hilbert2007}
{Hilbert} S.,  {White} S.~D.~M.,  {Hartlap} J.,   {Schneider} P.,  2007,
  \mn@doi [\mnras] {10.1111/j.1365-2966.2007.12391.x}, \href
  {http://adsabs.harvard.edu/abs/2007MNRAS.382..121H} {382, 121}

\bibitem[\protect\citeauthoryear{{Hilbert}, {Hartlap}, {White}  \&
  {Schneider}}{{Hilbert} et~al.}{2009}]{Hilbert2009}
{Hilbert} S.,  {Hartlap} J.,  {White} S.~D.~M.,   {Schneider} P.,  2009,
  \mn@doi [\aap] {10.1051/0004-6361/200811054}, \href
  {http://adsabs.harvard.edu/abs/2009A%26A...499...31H} {499, 31}

\bibitem[\protect\citeauthoryear{{Hinshaw} et~al.,}{{Hinshaw}
  et~al.}{2013}]{Hinshaw2013}
{Hinshaw} G.,  et~al., 2013, \mn@doi [\apjs] {10.1088/0067-0049/208/2/19},
  \href {http://adsabs.harvard.edu/abs/2013ApJS..208...19H} {208, 19}

\bibitem[\protect\citeauthoryear{{Kochanek}}{{Kochanek}}{2002}]{Kochanek2002}
{Kochanek} C.~S.,  2002, \mn@doi [\apj] {10.1086/342476}, \href
  {http://adsabs.harvard.edu/abs/2002ApJ...578...25K} {578, 25}

\bibitem[\protect\citeauthoryear{{Kochanek}, {Morgan}, {Falco}, {McLeod},
  {Winn}, {Dembicky}  \& {Ketzeback}}{{Kochanek} et~al.}{2006}]{Kochanek2006}
{Kochanek} C.~S.,  {Morgan} N.~D.,  {Falco} E.~E.,  {McLeod} B.~A.,  {Winn}
  J.~N.,  {Dembicky} J.,   {Ketzeback} B.,  2006, \mn@doi [\apj]
  {10.1086/499766}, \href {http://adsabs.harvard.edu/abs/2006ApJ...640...47K}
  {640, 47}

\bibitem[\protect\citeauthoryear{{Lesgourgues} \& {Tram}}{{Lesgourgues} \&
  {Tram}}{2014}]{Lesgourgues2014}
{Lesgourgues} J.,  {Tram} T.,  2014, \mn@doi [\jcap]
  {10.1088/1475-7516/2014/09/032}, \href
  {http://adsabs.harvard.edu/abs/2014JCAP...09..032L} {9, 032}

\bibitem[\protect\citeauthoryear{{Lewis} \& {Bridle}}{{Lewis} \&
  {Bridle}}{2002}]{Lewis2002}
{Lewis} A.,  {Bridle} S.,  2002, \mn@doi [\prd] {10.1103/PhysRevD.66.103511},
  \href {http://adsabs.harvard.edu/abs/2002PhRvD..66j3511L} {66, 103511}

\bibitem[\protect\citeauthoryear{{Liao} et~al.,}{{Liao}
  et~al.}{2015}]{Liao2015}
{Liao} K.,  et~al., 2015, \mn@doi [\apj] {10.1088/0004-637X/800/1/11}, \href
  {http://adsabs.harvard.edu/abs/2015ApJ...800...11L} {800, 11}

\bibitem[\protect\citeauthoryear{{Linder}}{{Linder}}{2011}]{Linder2011}
{Linder} E.~V.,  2011, \mn@doi [\prd] {10.1103/PhysRevD.84.123529}, \href
  {http://adsabs.harvard.edu/abs/2011PhRvD..84l3529L} {84, 123529}

\bibitem[\protect\citeauthoryear{{Magain}, {Courbin}  \& {Sohy}}{{Magain}
  et~al.}{1998}]{Magain1998}
{Magain} P.,  {Courbin} F.,   {Sohy} S.,  1998, \mn@doi [\apj]
  {10.1086/305187}, \href {http://adsabs.harvard.edu/abs/1998ApJ...494..472M}
  {494, 472}

\bibitem[\protect\citeauthoryear{{Marshall}, {Rajguru}  \& {Slosar}}{{Marshall}
  et~al.}{2006}]{Marshall2006}
{Marshall} P.,  {Rajguru} N.,   {Slosar} A.,  2006, \mn@doi [\prd]
  {10.1103/PhysRevD.73.067302}, \href
  {http://adsabs.harvard.edu/abs/2006PhRvD..73f7302M} {73, 067302}

\bibitem[\protect\citeauthoryear{{McCully}, {Keeton}, {Wong}  \&
  {Zabludoff}}{{McCully} et~al.}{2014}]{McCully2014}
{McCully} C.,  {Keeton} C.~R.,  {Wong} K.~C.,   {Zabludoff} A.~I.,  2014,
  \mn@doi [\mnras] {10.1093/mnras/stu1316}, \href
  {http://adsabs.harvard.edu/abs/2014MNRAS.443.3631M} {443, 3631}

\bibitem[\protect\citeauthoryear{{McCully}, {Keeton}, {Wong}  \&
  {Zabludoff}}{{McCully} et~al.}{2016}]{McCully2016}
{McCully} C.,  {Keeton} C.~R.,  {Wong} K.~C.,   {Zabludoff} A.~I.,  2016,
  ArXiv1601.05417, \href {http://adsabs.harvard.edu/abs/2016arXiv160105417M} {}

\bibitem[\protect\citeauthoryear{Molinari, Durand  \& Sabatier}{Molinari
  et~al.}{2004}]{Molinari2004}
Molinari N.,  Durand J.,   Sabatier R.,  2004, \mn@doi [Computational
  Statistics {\&} Data Analysis] {10.1016/S0167-9473(02)00343-2}, 45, 159

\bibitem[\protect\citeauthoryear{{Momcheva}, {Williams}, {Cool}, {Keeton}  \&
  {Zabludoff}}{{Momcheva} et~al.}{2015}]{Momcheva2015}
{Momcheva} I.~G.,  {Williams} K.~A.,  {Cool} R.~J.,  {Keeton} C.~R.,
  {Zabludoff} A.~I.,  2015, \mn@doi [\apjs] {10.1088/0067-0049/219/2/29}, \href
  {http://adsabs.harvard.edu/abs/2015ApJS..219...29M} {219, 29}

\bibitem[\protect\citeauthoryear{{Morgan}, {Kochanek}, {Pevunova}  \&
  {Schechter}}{{Morgan} et~al.}{2005}]{Morgan2005}
{Morgan} N.~D.,  {Kochanek} C.~S.,  {Pevunova} O.,   {Schechter} P.~L.,  2005,
  \mn@doi [\aj] {10.1086/430145}, \href
  {http://adsabs.harvard.edu/abs/2005AJ....129.2531M} {129, 2531}

\bibitem[\protect\citeauthoryear{{Oguri} \& {Marshall}}{{Oguri} \&
  {Marshall}}{2010}]{O+M10}
{Oguri} M.,  {Marshall} P.~J.,  2010, \mn@doi [\mnras]
  {10.1111/j.1365-2966.2010.16639.x}, \href
  {http://adsabs.harvard.edu/abs/2010MNRAS.405.2579O} {405, 2579}

\bibitem[\protect\citeauthoryear{{Padmanabhan}, {Xu}, {Eisenstein}, {Scalzo},
  {Cuesta}, {Mehta}  \& {Kazin}}{{Padmanabhan} et~al.}{2012}]{Padmanabhan2012}
{Padmanabhan} N.,  {Xu} X.,  {Eisenstein} D.~J.,  {Scalzo} R.,  {Cuesta} A.~J.,
   {Mehta} K.~T.,   {Kazin} E.,  2012, \mn@doi [\mnras]
  {10.1111/j.1365-2966.2012.21888.x}, \href
  {http://adsabs.harvard.edu/abs/2012MNRAS.427.2132P} {427, 2132}

\bibitem[\protect\citeauthoryear{{Pelt}, {Kayser}, {Refsdal}  \&
  {Schramm}}{{Pelt} et~al.}{1996}]{Pelt1996}
{Pelt} J.,  {Kayser} R.,  {Refsdal} S.,   {Schramm} T.,  1996, \aap, \href
  {http://adsabs.harvard.edu/abs/1996A%26A...305...97P} {305, 97}

\bibitem[\protect\citeauthoryear{{Percival} et~al.,}{{Percival}
  et~al.}{2010}]{Percival2010}
{Percival} W.~J.,  et~al., 2010, \mn@doi [\mnras]
  {10.1111/j.1365-2966.2009.15812.x}, \href
  {http://adsabs.harvard.edu/abs/2010MNRAS.401.2148P} {401, 2148}

\bibitem[\protect\citeauthoryear{{Planck Collaboration} et~al.,}{{Planck
  Collaboration} et~al.}{2014}]{Planck2013}
{Planck Collaboration} et~al., 2014, \mn@doi [\aap]
  {10.1051/0004-6361/201321591}, \href
  {http://adsabs.harvard.edu/abs/2014A%26A...571A..16P} {571, A16}

\bibitem[\protect\citeauthoryear{{Planck Collaboration} et~al.,}{{Planck
  Collaboration} et~al.}{2015a}]{Planck2015}
{Planck Collaboration} et~al., 2015a, ArXiv1502.01589, \href
  {http://adsabs.harvard.edu/abs/2015arXiv150201589P} {}

\bibitem[\protect\citeauthoryear{{Planck Collaboration} et~al.,}{{Planck
  Collaboration} et~al.}{2015b}]{PlanckLensing2015}
{Planck Collaboration} et~al., 2015b, ArXiv1502.01591, \href
  {http://adsabs.harvard.edu/abs/2015arXiv150201591P} {}

\bibitem[\protect\citeauthoryear{{Planck Collaboration} et~al.,}{{Planck
  Collaboration} et~al.}{2015c}]{PlanckTopology2015}
{Planck Collaboration} et~al., 2015c, preprint, \href
  {http://adsabs.harvard.edu/abs/2015arXiv150201593P} {} (\mn@eprint {arXiv}
  {1502.01593})

\bibitem[\protect\citeauthoryear{Rathna~Kumar et~al.,}{Rathna~Kumar
  et~al.}{2013}]{Rathnakumar2013}
Rathna~Kumar S.,  et~al., 2013, A{\&}A, 557, A44

\bibitem[\protect\citeauthoryear{{Refsdal}}{{Refsdal}}{1964}]{Refsdal1964}
{Refsdal} S.,  1964, \mnras, \href
  {http://adsabs.harvard.edu/cgi-bin/nph-bib_query?bibcode=1964MNRAS.128..307R&db_key=AST}
  {128, 307}

\bibitem[\protect\citeauthoryear{{Riess} et~al.,}{{Riess}
  et~al.}{2016}]{Riess2016}
{Riess} A.~G.,  et~al., 2016, \mn@doi [\apj] {10.3847/0004-637X/826/1/56},
  \href {http://adsabs.harvard.edu/abs/2016ApJ...826...56R} {826, 56}

\bibitem[\protect\citeauthoryear{{Rigault} et~al.,}{{Rigault}
  et~al.}{2015}]{Rigault2015}
{Rigault} M.,  et~al., 2015, \mn@doi [\apj] {10.1088/0004-637X/802/1/20}, \href
  {http://adsabs.harvard.edu/abs/2015ApJ...802...20R} {802, 20}

\bibitem[\protect\citeauthoryear{{Salvatelli}, {Marchini}, {Lopez-Honorez}  \&
  {Mena}}{{Salvatelli} et~al.}{2013}]{Salvatelli2013}
{Salvatelli} V.,  {Marchini} A.,  {Lopez-Honorez} L.,   {Mena} O.,  2013,
  \mn@doi [\prd] {10.1103/PhysRevD.88.023531}, \href
  {http://adsabs.harvard.edu/abs/2013PhRvD..88b3531S} {88, 023531}

\bibitem[\protect\citeauthoryear{{Schneider} \& {Sluse}}{{Schneider} \&
  {Sluse}}{2013}]{Schneider2013}
{Schneider} P.,  {Sluse} D.,  2013, \mn@doi [\aap]
  {10.1051/0004-6361/201321882}, \href
  {http://adsabs.harvard.edu/abs/2013A%26A...559A..37S} {559, A37}

\bibitem[\protect\citeauthoryear{{Schneider} \& {Sluse}}{{Schneider} \&
  {Sluse}}{2014}]{Schneider2014}
{Schneider} P.,  {Sluse} D.,  2014, \mn@doi [\aap]
  {10.1051/0004-6361/201322106}, \href
  {http://adsabs.harvard.edu/abs/2014A%26A...564A.103S} {564, A103}

\bibitem[\protect\citeauthoryear{{Sluse} \& {Tewes}}{{Sluse} \&
  {Tewes}}{2014}]{Sluse2014}
{Sluse} D.,  {Tewes} M.,  2014, \mn@doi [\aap] {10.1051/0004-6361/201424776},
  \href {http://adsabs.harvard.edu/abs/2014A%26A...571A..60S} {571, A60}

\bibitem[\protect\citeauthoryear{{Sluse}, {Hutsem{\'e}kers}, {Courbin},
  {Meylan}  \& {Wambsganss}}{{Sluse} et~al.}{2012}]{Sluse2012}
{Sluse} D.,  {Hutsem{\'e}kers} D.,  {Courbin} F.,  {Meylan} G.,   {Wambsganss}
  J.,  2012, \mn@doi [\aap] {10.1051/0004-6361/201219125}, \href
  {http://adsabs.harvard.edu/abs/2012A%26A...544A..62S} {544, A62}

\bibitem[\protect\citeauthoryear{{Sorce}, {Tully}  \& {Courtois}}{{Sorce}
  et~al.}{2012}]{Sorce2012}
{Sorce} J.~G.,  {Tully} R.~B.,   {Courtois} H.~M.,  2012, \mn@doi [\apjl]
  {10.1088/2041-8205/758/1/L12}, \href
  {http://adsabs.harvard.edu/abs/2012ApJ...758L..12S} {758, L12}

\bibitem[\protect\citeauthoryear{{Spergel}, {Flauger}  \& {Hlo{\v
  z}ek}}{{Spergel} et~al.}{2015}]{Spergel2015}
{Spergel} D.~N.,  {Flauger} R.,   {Hlo{\v z}ek} R.,  2015, \mn@doi [\prd]
  {10.1103/PhysRevD.91.023518}, \href
  {http://adsabs.harvard.edu/abs/2015PhRvD..91b3518S} {91, 023518}

\bibitem[\protect\citeauthoryear{{Suyu} \& {Halkola}}{{Suyu} \&
  {Halkola}}{2010}]{Suyu2010a}
{Suyu} S.~H.,  {Halkola} A.,  2010, \mn@doi [\aap]
  {10.1051/0004-6361/201015481}, \href
  {http://adsabs.harvard.edu/abs/2010A%26A...524A..94S} {524, A94}

\bibitem[\protect\citeauthoryear{{Suyu}, {Marshall}, {Hobson}  \&
  {Blandford}}{{Suyu} et~al.}{2006}]{Suyu2006}
{Suyu} S.~H.,  {Marshall} P.~J.,  {Hobson} M.~P.,   {Blandford} R.~D.,  2006,
  \mn@doi [\mnras] {10.1111/j.1365-2966.2006.10733.x}, \href
  {http://adsabs.harvard.edu/cgi-bin/nph-bib_query?bibcode=2006MNRAS.3
  71..983S&db_key=AST} {371, 983}

\bibitem[\protect\citeauthoryear{{Suyu}, {Marshall}, {Auger}, {Hilbert},
  {Blandford}, {Koopmans}, {Fassnacht}  \& {Treu}}{{Suyu}
  et~al.}{2010}]{Suyu2010b}
{Suyu} S.~H.,  {Marshall} P.~J.,  {Auger} M.~W.,  {Hilbert} S.,  {Blandford}
  R.~D.,  {Koopmans} L.~V.~E.,  {Fassnacht} C.~D.,   {Treu} T.,  2010, \mn@doi
  [\apj] {10.1088/0004-637X/711/1/201}, \href
  {http://adsabs.harvard.edu/abs/2010ApJ...711..201S} {711, 201}

\bibitem[\protect\citeauthoryear{{Suyu} et~al.,}{{Suyu}
  et~al.}{2012}]{Suyu2012}
{Suyu} S.~H.,  et~al., 2012, \mn@doi [\apj] {10.1088/0004-637X/750/1/10}, \href
  {http://adsabs.harvard.edu/abs/2012ApJ...750...10S} {750, 10}

\bibitem[\protect\citeauthoryear{{Suyu} et~al.,}{{Suyu}
  et~al.}{2013}]{Suyu2013}
{Suyu} S.~H.,  et~al., 2013, \mn@doi [\apj] {10.1088/0004-637X/766/2/70}, \href
  {http://adsabs.harvard.edu/abs/2013ApJ...766...70S} {766, 70}

\bibitem[\protect\citeauthoryear{{Suyu} et~al.,}{{Suyu}
  et~al.}{2014}]{Suyu2014}
{Suyu} S.~H.,  et~al., 2014, \mn@doi [\apjl] {10.1088/2041-8205/788/2/L35},
  \href {http://adsabs.harvard.edu/abs/2014ApJ...788L..35S} {788, L35}

\bibitem[\protect\citeauthoryear{{Tewes}, {Courbin}  \& {Meylan}}{{Tewes}
  et~al.}{2013a}]{Tewes2013a}
{Tewes} M.,  {Courbin} F.,   {Meylan} G.,  2013a, \mn@doi [\aap]
  {10.1051/0004-6361/201220123}, \href
  {http://adsabs.harvard.edu/abs/2013A%26A...553A.120T} {553, A120}

\bibitem[\protect\citeauthoryear{{Tewes} et~al.,}{{Tewes}
  et~al.}{2013b}]{Tewes2013b}
{Tewes} M.,  et~al., 2013b, A{\&}A, 556, A22

\bibitem[\protect\citeauthoryear{{Treu}}{{Treu}}{2010}]{Treu2010}
{Treu} T.,  2010, \mn@doi [\araa] {10.1146/annurev-astro-081309-130924}, \href
  {http://adsabs.harvard.edu/abs/2010ARA%26A..48...87T} {48, 87}

\bibitem[\protect\citeauthoryear{{Treu} \& {Marshall}}{{Treu} \&
  {Marshall}}{2016}]{Treu2016}
{Treu} T.,  {Marshall} P.~J.,  2016, \aapr, in press, \href
  {http://adsabs.harvard.edu/abs/2016arXiv160505333T} {}

\bibitem[\protect\citeauthoryear{{Unruh}, {Schneider}  \& {Sluse}}{{Unruh}
  et~al.}{2016}]{Unruh2016}
{Unruh} S.,  {Schneider} P.,   {Sluse} D.,  2016, ArXiv1606.04321, \href
  {http://adsabs.harvard.edu/abs/2016arXiv160604321U} {}

\bibitem[\protect\citeauthoryear{{Weinberg}, {Mortonson}, {Eisenstein},
  {Hirata}, {Riess}  \& {Rozo}}{{Weinberg} et~al.}{2013}]{Weinberg2013}
{Weinberg} D.~H.,  {Mortonson} M.~J.,  {Eisenstein} D.~J.,  {Hirata} C.,
  {Riess} A.~G.,   {Rozo} E.,  2013, \mn@doi [\physrep]
  {10.1016/j.physrep.2013.05.001}, \href
  {http://adsabs.harvard.edu/abs/2013PhR...530...87W} {530, 87}

\bibitem[\protect\citeauthoryear{{Wisotzki}, {Christlieb}, {Bade}, {Beckmann},
  {K{\"o}hler}, {Vanelle}  \& {Reimers}}{{Wisotzki}
  et~al.}{2000}]{Wisotzki2000}
{Wisotzki} L.,  {Christlieb} N.,  {Bade} N.,  {Beckmann} V.,  {K{\"o}hler} T.,
  {Vanelle} C.,   {Reimers} D.,  2000, \aap, \href
  {http://adsabs.harvard.edu/abs/2000A%26A...358...77W} {358, 77}

\bibitem[\protect\citeauthoryear{{Wisotzki}, {Schechter}, {Bradt},
  {Heinm{\"u}ller}  \& {Reimers}}{{Wisotzki} et~al.}{2002}]{Wisotzki2002}
{Wisotzki} L.,  {Schechter} P.~L.,  {Bradt} H.~V.,  {Heinm{\"u}ller} J.,
  {Reimers} D.,  2002, \mn@doi [\aap] {10.1051/0004-6361:20021213}, \href
  {http://adsabs.harvard.edu/abs/20@article{Molinari2004, author = {Nicolas
  Molinari and Jean{-}Fran{\c{c}}ois Durand and Robert Sabatier}, title =
  {Bounded optimal knots for regression splines}, journal = {Computational
  Statistics {\&} Data Analysis}, volume = {45}, number = {2}, pages =
  {159--178}, year = {2004}, url =
  {http://dx.doi.org/10.1016/S0167-9473(02)00343-2}, doi =
  {10.1016/S0167-9473(02)00343-2}, timestamp = {Wed, 18 Apr 2007 14:41:36
  +0200}, biburl =
  {http://dblp.uni-trier.de/rec/bib/journals/csda/MolinariDS04}, bibsource =
  {dblp computer science bibliography, http://dblp.org} }02A%26A...395...17W}
  {395, 17}

\bibitem[\protect\citeauthoryear{{Wong}, {Keeton}, {Williams}, {Momcheva}  \&
  {Zabludoff}}{{Wong} et~al.}{2011}]{Wong2011}
{Wong} K.~C.,  {Keeton} C.~R.,  {Williams} K.~A.,  {Momcheva} I.~G.,
  {Zabludoff} A.~I.,  2011, \mn@doi [\apj] {10.1088/0004-637X/726/2/84}, \href
  {http://adsabs.harvard.edu/abs/2011ApJ...726...84W} {726, 84}

\bibitem[\protect\citeauthoryear{{Wucknitz}}{{Wucknitz}}{2002}]{Wucknitz2002}
{Wucknitz} O.,  2002, \mn@doi [\mnras] {10.1046/j.1365-8711.2002.05426.x},
  \href {http://adsabs.harvard.edu/abs/2002MNRAS.332..951W} {332, 951}

\bibitem[\protect\citeauthoryear{{Xu}, {Sluse}, {Schneider}, {Springel},
  {Vogelsberger}, {Nelson}  \& {Hernquist}}{{Xu} et~al.}{2016}]{Xu2016}
{Xu} D.,  {Sluse} D.,  {Schneider} P.,  {Springel} V.,  {Vogelsberger} M.,
  {Nelson} D.,   {Hernquist} L.,  2016, \mn@doi [\mnras]
  {10.1093/mnras/stv2708}, \href
  {http://adsabs.harvard.edu/abs/2016MNRAS.456..739X} {456, 739}

\makeatother
\end{thebibliography}
\bibliographystyle{mnras}

%-------------------------------------------------------------------------------

\label{lastpage}
\end{document}